\documentclass[twocolumn]{aastex63}

\usepackage{lipsum, babel}
\usepackage{footmisc} % restores normal footnote commands

\newcommand\tess{TESS}
\newcommand\gaia{\textit{Gaia}}

\newcommand{\unit}[1]{\ensuremath{\, \mathrm{#1}}} %Remove if necessary

\submitjournal{ApJ}

\begin{document}

\title{Discovery of a Nearby Habitable Zone Super-Earth Candidate Amenable to Direct Imaging}

\author[0000-0001-7708-2364]{Corey Beard}
\altaffiliation{NASA FINESST Fellow}
\affiliation{Department of Physics \& Astronomy, The University of California, Irvine, Irvine, CA 92697, USA}

\author[0000-0003-0149-9678]{Paul Robertson}
\affiliation{Department of Physics \& Astronomy, The University of California, Irvine, Irvine, CA 92697, USA}

\author[0000-0001-8342-7736]{Jack Lubin}
\affil{Department of Physics \& Astronomy, University of California Los Angeles, Los Angeles, CA 90095, USA}
%confirmed

%%%%%%%%%%%%%%%%%%%%%%%%%%

\author[0000-0001-6545-639X]{Eric B.\ Ford}
\affil{Department of Astronomy \& Astrophysics, 525 Davey Laboratory, 251 Pollock Road, Penn State, University Park, PA, 16802, USA}
\affil{Center for Exoplanets and Habitable Worlds, 525 Davey Laboratory, 251 Pollock Road, Penn State, University Park, PA, 16802, USA}
\affil{Institute for Computational and Data Sciences, Penn State, University Park, PA, 16802, USA}
\affil{Center for Astrostatistics, 525 Davey Laboratory, 251 Pollock Road, Penn State, University Park, PA, 16802, USA}
%confirmed

\author[0000-0001-9596-7983]{Suvrath Mahadevan}
\affil{Department of Astronomy \& Astrophysics, 525 Davey Laboratory, 251 Pollock Road, Penn State, University Park, PA, 16802, USA}
\affil{Center for Exoplanets and Habitable Worlds, 525 Davey Laboratory, 251 Pollock Road, Penn State, University Park, PA, 16802, USA}
\affil{Astrobiology Research Center, 525 Davey Laboratory, 251 Pollock Road, Penn State, University Park, PA, 16802, USA} 
%confirmed

\author[0000-0001-7409-5688]{Gudmundur Stefansson}
\affil{Anton Pannekoek Institute for Astronomy, 904 Science Park, University of Amsterdam, Amsterdam, 1098 XH}
%confirmed

\author[0000-0001-6160-5888]{Jason T.\ Wright}
\affil{Department of Astronomy \& Astrophysics, 525 Davey Laboratory, 251 Pollock Road, Penn State, University Park, PA, 16802, USA}
\affil{Center for Exoplanets and Habitable Worlds, 525 Davey Laboratory, 251 Pollock Road, Penn State, University Park, PA, 16802, USA}
\affil{Penn State Extraterrestrial Intelligence Center, 525 Davey Laboratory, 251 Pollock Road, Penn State, University Park, PA, 16802, USA}
%confirmed

%%%%%%%%%%%%%%%%%%%%%%%%%%

\author[0000-0002-7188-1648]{Eric Wolf}
\affil{Laboratory for Atmospheric and Space Physics, Department of Atmospheric and Oceanic Sciences, University of Colorado, Boulder, CO, USA}
%confirmed

\author[0000-0002-5060-1993]{Vincent Kofman}
\affil{NASA Goddard Space Flight Center, Greenbelt, MD 20771, USA}
\affil{Department of Physics, American University, Washington, DC, 20016, USA}
%confirmed

\author[0000-0002-5638-4344]{Vidya Venkatesan}
\altaffiliation{NASA FINESST Fellow}
\affiliation{Department of Physics \& Astronomy, The University of California, Irvine, Irvine, CA 92697, USA}
%confirmed

\author[0000-0002-5893-2471]{Ravi Kopparapu}
\affil{NASA Goddard Space Flight Center, Greenbelt, MD 20771, USA}
%confirmed

%%%%%%%%%%%%%%%%%%%%%

\author[0009-0001-6669-264X]{Roan Arendtsz}
\affil{Carleton College, One North College St., Northfield, MN 55057, USA}

\author[0000-0002-5034-9476]{Rae Holcomb}
\affil{Department of Physics \& Astronomy, The University of California, Irvine, Irvine, CA 92697, USA}
%confirmed

\author[0000-0001-6301-896X]{Raquel A. Martinez}
\affil{Department of Physics and Biophysics, University of San Diego, 5998 Alcal\'a Park, San Diego, CA 92110}
%confirmed

\author[0000-0001-6871-6775]{Stephanie Sallum}
\affil{Department of Astronomy and Astrophysics, UC Santa Cruz, 1156 High St. Santa Cruz, CA 95064}
%confirmed

\author[0000-0002-4927-9925]{Jacob K. Luhn}
\altaffiliation{NASA Postdoctoral Fellow}
\affil{Jet Propulsion Laboratory, California Institute of Technology, 4800 Oak Grove Drive, Pasadena, CA 91109}
%confirmed

%%%%%%%%%%%%%%%%%%%%%

\author[0000-0003-4384-7220]{Chad F.\ Bender}
\affil{Steward Observatory, University of Arizona, 933 N.\ Cherry Ave, Tucson, AZ 85721, USA}
%confirmed

\author[0000-0002-6096-1749]{Cullen H.\ Blake}
\affil{Department of Physics and Astronomy, University of Pennsylvania, 209 S 33rd St, Philadelphia, PA 19104, USA}
%confirmed

\author[0000-0001-9662-3496]{William D. Cochran}
\affiliation{McDonald Observatory and Center for Planetary Systems Habitability, The University of Texas, Austin TX 78712 USA.}
%confirmed

\author[0000-0003-1439-2781]{Megan Delamer}
\affil{Department of Astronomy \& Astrophysics, 525 Davey Laboratory, 251 Pollock Road, Penn State, University Park, PA, 16802, USA}
\affil{Center for Exoplanets and Habitable Worlds, 525 Davey Laboratory, 251 Pollock Road, Penn State, University Park, PA, 16802, USA}
%confirmed

\author[0000-0002-2144-0764]{Scott A.\ Diddams}
\affil{Electrical, Computer \& Energy Engineering, 425 UCB University of Colorado, Boulder, CO 80309, USA}
\affil{Department of Physics, 390 UCB, University of Colorado, Boulder, CO 80309, USA}
%confirmed

\author[0000-0002-7714-6310]{Michael Endl}
\affiliation{Center for Planetary Systems Habitability, The University of Texas at Austin, Austin, TX 78712, USA}
\affiliation{Department of Astronomy, The University of Texas at Austin, TX, 78712, USA}
%confirmed

\author[0000-0003-1312-9391]{Samuel Halverson}
\affil{Jet Propulsion Laboratory, California Institute of Technology, 4800 Oak Grove Drive, Pasadena, California 91109}
%confirmed

\author[0000-0001-8401-4300]{Shubham Kanodia}
\affil{Earth and Planets Laboratory, Carnegie Science, 5241 Broad Branch Road, NW, Washington, DC 20015, USA}
%confirmed

\author[0000-0001-9626-0613]{Daniel M.\ Krolikowski}
\affil{Steward Observatory, University of Arizona, 933 N.\ Cherry Ave, Tucson, AZ 85721, USA}
%confirmed

\author[0000-0002-9082-6337]{Andrea S.J.\ Lin}
\affil{Department of Astronomy, California Institute of Technology, 1200 E California Blvd, Pasadena, CA 91125, USA}
%confirmed

\author[0000-0002-9632-9382]{Sarah E.\ Logsdon}
\affil{U.S. National Science Foundation National Optical-Infrared Astronomy Research Laboratory, 950 N.\ Cherry Ave., Tucson, AZ 85719, USA}
%confirmed

\author[0000-0003-0241-8956]{Michael W.\ McElwain}
\affil{Exoplanets and Stellar Astrophysics Laboratory, NASA Goddard Space Flight Center, Greenbelt, MD 20771, USA} 
%confirmed

\author[0000-0002-0048-2586]{Andrew Monson}
\affil{Steward Observatory, University of Arizona, 933 N.\ Cherry Ave, Tucson, AZ 85721, USA}
%confirmed

\author[0000-0001-8720-5612]{Joe P.\ Ninan}
\affil{Department of Astronomy and Astrophysics, Tata Institute of Fundamental Research, Homi Bhabha Road, Colaba, Mumbai 400005, India}
%confirmed

\author[0000-0002-2488-7123]{Jayadev Rajagopal}
\affil{U.S. National Science Foundation National Optical-Infrared Astronomy Research Laboratory, 950 N. Cherry Ave., Tucson, AZ 85719, USA}
%confirmed

\author[0000-0001-8127-5775]{Arpita Roy}
\affil{Astrophysics and Space Institute, Schmidt Sciences, New York, NY 10011, USA}
%confirmed

\author[0000-0002-4046-987X]{Christian Schwab}
\affil{School of Mathematical and Physical Sciences, Macquarie University, Balaclava Road, North Ryde, NSW 2109, Australia}
%confirmed

\author[0000-0002-4788-8858]{Ryan C. Terrien}
\affil{Carleton College, One North College St., Northfield, MN 55057, USA}
%confirmed

\correspondingauthor{Corey Beard}
\email{ccbeard@uci.edu}

\begin{abstract}

We present the discovery of GJ 251 c, a candidate super-Earth orbiting in the Habitable Zone (HZ) of its M dwarf host star. Using high-precision Habitable-zone Planet Finder (HPF) and NEID RVs, in conjunction with archival RVs from the Keck I High Resolution Echelle Spectrometer (HIRES),  the Calar Alto high-Resolution search for M dwarfs with Exoearths with Near-infrared and optical Echelle Spectrograph (CARMENES), and the SPectropolarimètre InfraROUge (SPIRou), we improve the measured parameters of the known planet, GJ 251 b ($P_{b}$ = 14.2370 days; $m \sin(i)$ = 3.85$^{+0.35}_{-0.33}$ M$_{\oplus}$), and we significantly constrain the minimum mass of GJ 251 c, placing it in a plausibly terrestrial regime (P$_{c}$ = 53.647 $\pm$ 0.044 days; $ m \sin i_{c}$ = 3.84 $\pm$ 0.75 M$_{\oplus}$). Using activity mitigation techniques that leverage chromatic information content, we perform a color-dependent analysis of the system and a detailed comparison of more than 50 models that describe the nature of the planets and stellar activity in the system. Due to GJ 251's proximity to Earth (5.5 pc), next generation, thirty meter class telescopes will likely be able to image terrestrial planets in GJ 251's HZ. In fact, GJ 251 c is currently the best candidate for terrestrial, HZ planet imaging in the Northern Sky.

\end{abstract}


\section{Introduction} \label{sec:intro}

Following recent discoveries from the Transiting Exoplanet Survey Satellite \citep[TESS;][]{ricker15}, the exoplanet community has reached a milestone: over 5000 exoplanets confirmed, with thousands of additional candidates likely to be validated in the near future. With the successful launch of JWST \citep{gardner06}, we have begun to deeply characterize the atmospheres of a variety of transiting planets \citep{lim24,damiano24}. Unfortunately, transmission spectroscopy is limited to a small subset of exoplanets that transit. Other methods, such as direct imaging, will be required to probe atmospheric information from non-transiting exoplanets \citep{marois08, currie23}. Instrumental limitations have historically restricted direct imaging to large, long period planets orbiting young stars in close proximity to Earth. Advances and future ground-based/space-based observatories and missions \citep[i.e. TMT; GMT; E-ELT; HWO; LIFE;][]{skidmore15, johns12, gilmozzi07, decadal20, quanz22} will likely allow the next generation of imagers to push our direct imaging targets to further, dimmer systems with smaller planet-star angular separations and contrasts.

The concept of the Habitable Zone \citep[HZ;][]{kasting93,kopparapu13, kopparapu14, kane16} is used to highlight regions of orbital parameter space where temperate surface conditions of a planet are possible, though it does not guarantee habitability. While many planets are known to orbit within their star's HZ, few of these transit, and those that transit are often difficult to follow-up with other forms of observation \citep[e.g. radial velocities, direct imaging, etc.][]{hill22}. This can be a critical problem, as a precise mass can be crucial to the correct interpretation of atmospheric composition measurements \citep{batalha19}, a required next step to determine habitability or to confirm the presence of biosignatures.

If we wish to find nearby exoplanets in the HZ, transit surveys are not sufficient: the low probability for a planet to transit from Earth results in many missed, non-transiting exoplanets that can still be identified with RVs. In fact, it is expected that the closest HZ planets will be discovered via RV, not transits \citep{crass21,hardegree-ullman23}. Limitations on our ability to resolve planets remain, though this can be mitigated by studying nearby planets. A planet orbiting a nearby star will have a larger angular separation than the same planet orbiting a distant star, making direct observation easier. Closer stars also appear brighter, and thus are more amenable to many other forms of study.

GJ 251 is an early M dwarf and the 74$^{\rm{th}}$ closest star system to the Sun. GJ 251 is bright (V = 9.9 $\pm$ 0.1) and close (d = 5.58 $\pm$ 0.01 pc), making it an excellent target for future direct imaging missions. \cite{stock20} and \cite{ouldelhkim23} (hereafter S20 and OE23) each performed an analysis using CARMENES and SPIRou RV data, identifying a planet, GJ 251 b (P$_{b}$ = 14.2 days; $m \sin(i)$ = 3.85$^{+0.35}_{-0.33}$ M$_{\oplus}$). 

We present a newly discovered planet candidate, GJ 251 c. The candidate has a measured minimum mass of 3.84 $\pm$ 0.75 M$_{\oplus}$, making it plausibly terrestrial\footnote{Here we call a planet plausibly terrestrial if $m\sin(i)$ $<$ 5.0 M$_{\oplus}$ \citep{parc24}}. Due to its host star's proximity to Earth, GJ 251 c falls in a very narrow range of parameter space wherein a terrestrial, HZ exoplanet may be directly imaged via reflected light with the upcoming next-generation of extremely large telescopes (ELT).

GJ 251 has over 20 years of archival RVs, first taken with the High Resolution Echelle Spectrometer at Keck Observatory \citep[HIRES;][]{vogt94}, and with more recent observations taken by the Calar Alto high-Resolution search for M dwarfs with Exoearths with Near-infrared and optical Echelle Spectrograph \citep[CARMENES;][]{quirrenbach14}, and the SPectropolarimètre InfraROUge \citep[SPIRou;][]{donati18}. We combine these archival RVs with new RVs taken by the Habitable-zone Planet Finder \citep[HPF;][]{mahadevan12,mahadevan14}, and NEID \citep{schwab16} spectrometers.

Leveraging chromatic Gaussian processes kernels \citep{cale21} in conjunction with precise HPF and NEID RVs, we improve the measured parameters of GJ 251 b as well as identify a candidate super Earth, GJ 251 c. 

In \S \ref{sec:observations}, we give a summary of the data used in our analysis. In \S \ref{sec:stellar}, we detail our estimation of the system's stellar parameters. In \S \ref{sec:periodogram analysis}, we specify the steps taken to identify additional planets in the system, and in \S \ref{sec:mcmc_nested} the analysis methods used when considering the data, and the model comparisons that ultimately led to our conclusion. In \S \ref{sec:interpret}, we present our interpretation of the various signals in the dataset. In \S \ref{sec:other_data}, we look for astrometric and transit signatures in non-RV datasets. In \S \ref{sec:discussion}, we discuss our findings, and the implications for the system. Finally, \S \ref{sec:summary} summarizes our results and conclusions.

\section{Observations}
\label{sec:observations}

\subsection{RVs with Keck/HIRES}
\label{sec:hires}

High precision RVs of GJ 251 were obtained using the HIRES spectrograph \citep{vogt94} mounted on the 10 m Keck-I telescope located in Hawaii, USA. A total of 78 HIRES RVs were obtained between 1997 December 1 and 2019 November 9, taken over 64 observing nights. We use RVs derived from \cite{butler17}, which are reduced using the iodine-cell method described in \cite{butler96}. HIRES' wavelength coverage is 320-800 nm (though the iodine cell only covers 500-600 nm).

The HIRES spectrograph was upgraded in August 2004, resulting in an offset between RVs measured before and after this upgrade. GJ 251's HIRES observations lie on both sides of this upgrade. We label any RVs taken before August 2004 as ``HIRES-pre" (N$_{pre}$ = 14), and RVs taken after as ``HIRES-post" (N$_{post}$ = 64). For purposes of data analysis, we treat these as separate instruments, with different linear offset and jitter terms.

GJ 251 HIRES observations have a median exposure time of 500 seconds, and a median S/N of 186 at 550 nm. The median 1$\sigma$ RV uncertainty is 2.12 m s$^{-1}$.

\subsection{RVs with the CARMENES Spectrograph}
\label{sec:carmenes}

We utilized the high precision CARMENES RVs published in S20 and \cite{ribas23}. These datasets collectively contain 265 RVs of GJ 251 over 258 observing nights, starting on 2016 January 8, and ending on 2020 December 29. CARMENES is a double-channel \'{e}chelle spectrograph located at the Calar Alto Observatory in Almer\'{i}a, Spain \citep{quirrenbach14}. The RVs used in our analysis were extracted from the visible arm of the CARMENES spectrograph (550-1700 nm), as the Near-Infrared (NIR) arm exhibited a systematic scatter on the order of a few m s$^{-1}$. CARMENES RVs were extracted using the \texttt{SpEctrum Radial Velocity AnaLyser} pipeline \citep[\texttt{SERVAL};][]{zechmeister18}. RVs were corrected for a nightly zero-point offset (NZP) in \cite{ribas23} prior to analysis.

The CARMENES RVs of GJ 251 have a median exposure time of 509 s, and a median S/N of 134 at 746 nm. The resulting RVs have a median 1$\sigma$ errorbar of 1.25 m s$^{-1}$.

\subsection{RVs with SPIRou}
\label{sec:spirou}

We utilize an additional 177 RV observations of GJ 251 taken using the SPIRou spectropolarimeter, located at the Canada-France-Hawaii Telescope (CFHT) at Mauna Kea, Hawaii \citep{donati20}. SPIRou is an infrared spectropolarimeter, with a broad wavelength range of 980-2440 nm. 

OE23 reduced raw SPIRou spectra using the SPIRou data reduction pipeline \citep{cook22}. OE23 additionally cleaned the data of systematics using the line-by-line method \citep[LBL;][]{artigau22,dumusque18} implemented with the \texttt{WAPITI} software package (OE23). Tellurics are removed by a hybrid model-based and empirical correction in the reduction \citep{artigau22}.

SPIRou RV data have a median RV error of 1.76 m s$^{-1}$ and a total observing baseline of 1212 days. 

\subsection{RVs with the Habitable-Zone Planet Finder}\label{sec:hpfrvs}
\label{sec:hpf}

We observed GJ 251 using with the Habitable-zone Planet Finder \citep[HPF;][]{mahadevan12, mahadevan14}, a near-infrared, (\(808-1278\)\ nm), high precision RV spectrograph. HPF is located at the 10 m Hobby-Eberly Telescope (HET) at McDonald Observatory, Texas. HET is a fixed-altitude telescope with a roving pupil design. HPF is fiber-fed, with separate science, sky and simultaneous calibration fibers \citep{kanodia18a}, and has precise, milliKelvin-level thermal stability \citep{stefansson16}. 

HPF RVs were generated using a version of SERVAL adapted for use on HPF data as described in \cite[][the latter of which made the code public\footnote{\url{https://github.com/gummiks/hpfserval_lhs3154}}]{stefansson20_a, stefansson23}.

HPF required maintenance in May 2022, which necessitated a warm-up of the spectrograph's environmental vacuum chamber. Consequently, RVs taken before and after the break are subject to a different instrument offset. Similar to HIRES in \S \ref{sec:hires}, we treat the pre- and post-servicing HPF RVs as two separate instruments in our fits (N$_{pre}$ = 55; N$_{post}$  = 14).

We obtained 375 HPF observations over the course of 69 observing nights. Our HPF observations ranged from 16 December 2018 to 1 February 2024. HPF observations were typically carried out in multishot-sequences, ranging from 3-7 in a night. The median exposure time was 330 s per exposure, with a median S/N of 414 at the 18$^{th}$ wavelength order (1070 nm). This corresponds to a median unbinned error of 2.95 m s$^{-1}$ and a median nightly binned error of 1.23 m s$^{-1}$.

\subsection{RVs with the NEID Spectrometer}
\label{sec:neid}

We obtained RVs of GJ 251 using the high precision NEID spectrometer \citep{schwab16} on the WIYN 3.5\,m Telescope\footnote{The WIYN Observatory is a joint facility of the NSF's National Optical-Infrared Astronomy Research Laboratory, Indiana University, the University of Wisconsin-Madison, Pennsylvania State University, and Princeton University.} at Kitt Peak National Observatory (KPNO). We observed GJ 251 in High Resolution (HR) mode for best RV precision, with an average resolving power R = 110,000. NEID has a wavelength range between 380 and 930 nm.

We obtained 92 NEID RVs of GJ 251 over 52 observing nights between 2020 December 19 and 2024 January 18. We typically took two exposures of GJ 251 per night, with a median exposure time of 900 s per observation. This corresponds to a median S/N of 117 at 490 nm. The default NEID pipeline utilizes the Cross-Correlation Function \citep[CCF; ][]{baranne96} method to produce RVs. However, our analysis utilized NEID RVs estimated with \texttt{SERVAL} code that is adapted for use with NEID \citep{stefansson22}. Comparing the CCF-produced RVs of GJ 251 with the \texttt{SERVAL} produced RVs, the choice of pipeline is clear. NEID CCF RVs see a precision loss on later-type stars during reduction, most likely because the NEID CCF masks do not extend as far into long wavelengths as NEID's bandpass does. The median unbinned NEID CCF 1$\sigma$ errorbar is 2.6 m s$^{-1}$. \texttt{SERVAL} circumnavigates this problem, and produces extremely precise NEID RVs of GJ 251, with a median unbinned SERVAL 1$\sigma$ errorbar of 40 cm s$^{-1}$.

KPNO was threatened by a wildfire during the Summer of 2022, which required that NEID be put in a safe, standby mode during firefighting operations. Precise RV observations with NEID were possible starting in November 2022, though subsequent RVs are expected to have an instrumental offset due to vacuum breaking. Consequently, we treat pre-fire NEID RVs (N$_{pre}$ = 25) and post-fire NEID RVs (N$_{post}$ = 67) as separate instruments, consistent with HIRES and HPF in \S \ref{sec:hires} and \ref{sec:hpf}.

\subsubsection{Extracting Chromatic RVs}

Previous studies suggest that RVs may be less prone to certain kinds of activity contamination in the NIR \citep{crockett12,robertson20}. It is possible to generate less precise, but ``red" RVs from NEID by only utilizing photons from longer wavelengths. Such a precision loss would often disqualify such an approach from consideration, but GJ 251 is well suited to this method: GJ 251 is an early M Dwarf (M3), thus it is brightest in the NIR. Additionally, due to its proximity to Earth, it is particularly bright, allowing for the acquisition of high S/N spectra with minimal time investment.

\begin{figure*}
    \centering
    \includegraphics[width=\textwidth]{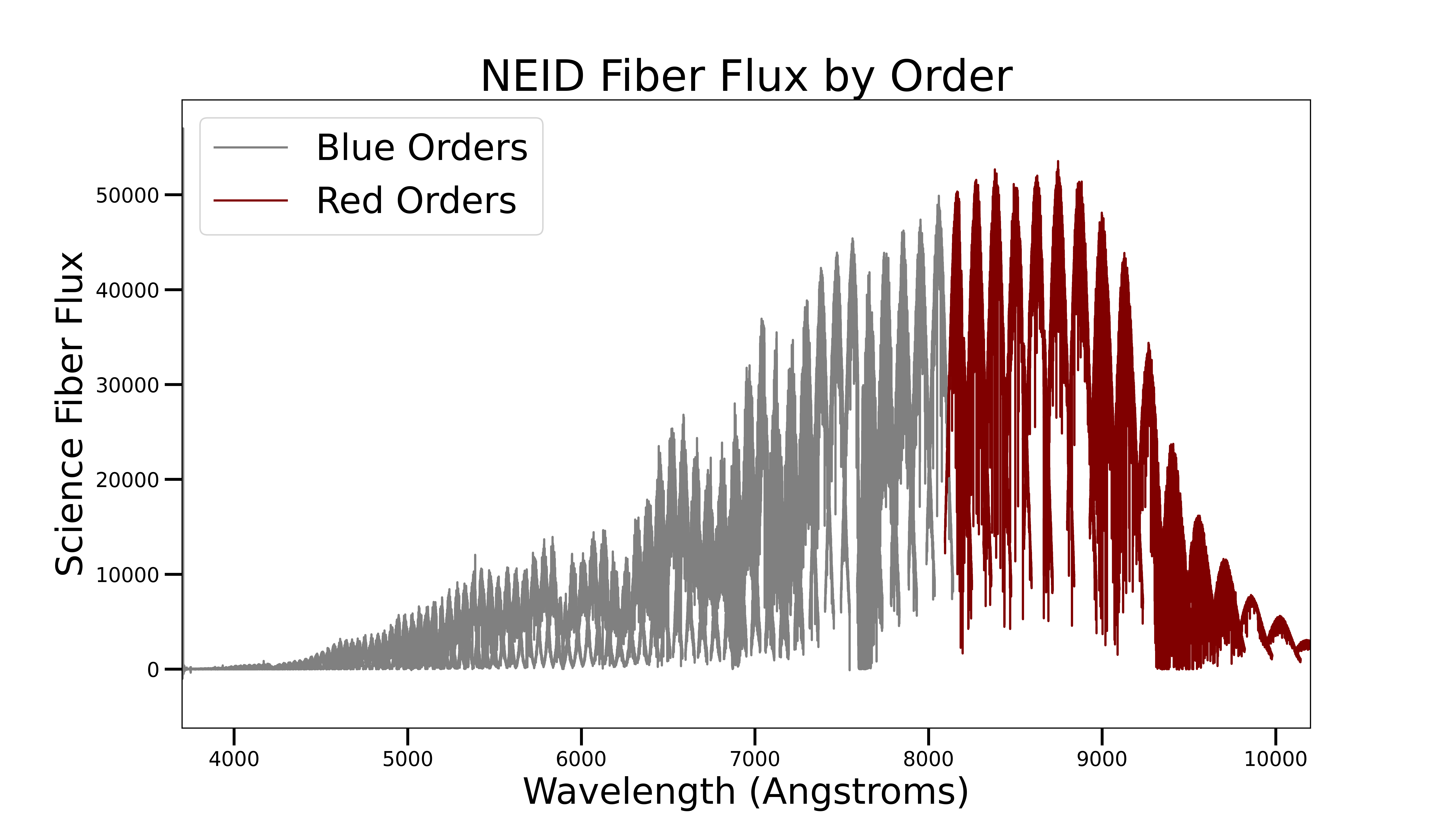}
    \caption{Science fiber flux as a function of wavelength for the GJ 251 NEID spectra taken on 2021 October 11. Each order of flux is centered around a different peak wavelength. The NEID spectrograph has 174 orders, though the instrument response is primarily meaningful for orders 52 through 174. Notable is the low S/N for the bluer orders. As an early M dwarf, GJ 251 contains most of its flux at longer wavelengths. The section highlighted in red corresponds to a peak order wavelength greater than 8000 \AA. These red orders are used in our creation of ``red" NEID RVs for GJ 251.}
    \label{fig:red_orders}
\end{figure*}

We experimented with utilizing different orders to create pseudo-red RVs, but ultimately decided that an 800 nm lower limit provided the best tradeoff between longer wavelength information content and high precision. Consequently, when creating pseudo-infrared NEID RVs, we only used NEID orders where the peak flux occurs at a wavelength greater than 800 nm. A plot of the by-order flux, and our red cutoffs, is visible in Figure \ref{fig:red_orders}. Using \texttt{SERVAL}, we first reduce the RV spectra by order, with an RV value and photon-limited uncertainty (combined with known instrumental error) assigned to each. To create psuedo infrared-RVs, we identified each order where the peak flux occurred at a wavelength $>$ 800 nm, and we took a weighted average combination of the by-order RVs. \texttt{SERVAL} produces uncertainties in each order during reduction, making this weighted combination a straightforward application after \texttt{SERVAL} reduction.

We establish here the nomenclature that will be used throughout the paper. NEID RVs utilizing orders 52-174, we will call ``white" NEID RVs, as they utilize all photon information regardless of color. On the other hand, we will refer to the RVs that have been specifically created from the reddest orders (peak flux occurring at wavelength $>$ 800 nm; orders 150-174) as ``red" NEID RVs. Additionally, the combined datasets (utilizing all instruments) will sometimes be referred to as the white dataset and the red dataset respectively, though the vast majority of the RVs in each dataset are identical (only the NEID RVs are different). Data analysis of GJ 251 was performed extensively with both white and red reduced NEID RVs.

\section{Stellar Parameters}\label{sec:stellar}

GJ 251 is included in the library of touchstone, high resolution spectra obtained with Keck/HIRES in \cite{yee17}, and so reliable literature stellar parameters are available. We additionally report our own measured stellar parameters from HPF high S/N spectra. 

We used \texttt{EXOFASTv2} \citep{eastman13} to model the spectral energy distribution (SED) of GJ 251, and to derive model-dependent constraints on the stellar mass, radius, and age. \texttt{EXOFASTv2} utilizes the BT-NextGen stellar atmospheric models \citep{allard12} during SED fits.  We used Gaussian priors for the estimate of 2MASS (\(JHK_{s}\)), Johnson (\(B V\)), and \textit{Wide-field Infrared Survey Explorer} magnitudes (WISE; $W1$, $W2$, $W3$, and $W4$) \citep[][]{wright10}. The library values of $T_{\rm{eff}}$, [Fe/H], and $\log g$ from \cite{yee17} were used as Gaussian priors during the SED fits, and the distance estimates from \cite{bailer-jones21} were used as priors for distance. Due to its proximity to Earth, we do not utilize a dust-extinction term in our fits. Our final model results are consistent with those derived in S20 and \cite{yee17}, and are visible in Table \ref{tab:stellartable}.

\begin{deluxetable*}{lccc}
\tablecaption{Summary of stellar parameters for GJ 251. \label{tab:stellartable}}
\tabletypesize{\scriptsize}
\tablehead{\colhead{~~~Parameter}&  \colhead{Description}&
\colhead{Value}&
\colhead{Reference}}
\startdata
\multicolumn{4}{l}{\hspace{-0.2cm} Main identifiers:}  \\
~~~GJ & Gliese-Jahreiss Nearby Stars & 251 & Gliese-Jahreiss \\
~~~HD & Henry-Draper Catalog & 265866 & Henry-Draper \\
~~~HIP & Hipparcos Catalog & 33226 & Hipparcos \\
~~~LHS & Luyten Half-Second & 1879 & Luyten \\
~~~TIC & TESS Input Catalog  & 68581262 & Stassun \\
~~~2MASS & \(\cdots\) & J06544902+3316058 & 2MASS  \\
~~~Gaia DR3 & \(\cdots\) & 939072613334579328 & Gaia DR3\\
\multicolumn{4}{l}{\hspace{-0.2cm} Equatorial Coordinates, Proper Motion and Spectral Type:} \\
~~~$\alpha_{\mathrm{ICRS2016}}$ &  Right Ascension (RA) & 103.70012922(7) & Gaia DR3\\
~~~$\delta_{\mathrm{ICRS2016}}$ &  Declination (Dec) & 33.26640801(8) & Gaia DR3\\
~~~$\mu_{\alpha}$ &  Proper motion (RA, \unit{mas/yr}) &  -726.67 $\pm$ 0.03 & Gaia DR3\\
~~~$\mu_{\delta}$ &  Proper motion (Dec, \unit{mas/yr}) & -398.10 $\pm$ 0.02 & Gaia DR3 \\
~~~$d$ &  Distance in pc  & 5.58 $\pm$ 0.01 & Bailer-Jones \\
\multicolumn{4}{l}{\hspace{-0.2cm} Optical and near-infrared magnitudes:}  \\
~~~$B$ & Johnson B mag & 11.7 $\pm$ 0.1 & APASS\\
~~~$V$ & Johnson V mag & 9.9 $\pm$ 0.1 & APASS\\
~~~$g^{\prime}$ &  Sloan $g^{\prime}$ mag  & $10.79\pm0.04$ & APASS\\
~~~$T$  & TESS magnitude &  7.648 $\pm$ 0.008 & Stassun \\
~~~$J$ & $J$ mag & 6.10 $\pm$ 0.02 & 2MASS\\
~~~$H$ & $H$ mag & 5.53 $\pm$ 0.02 & 2MASS\\
~~~$K_s$ & $K_s$ mag & 5.27 $\pm$ 0.02 & 2MASS\\
~~~$W1$ &  WISE1 mag & $5.1 \pm 0.2$ & WISE\\
~~~$W2$ &  WISE2 mag & $4.80 \pm 0.08$ & WISE\\
~~~$W3$ &  WISE3 mag & $4.92\pm0.02$ & WISE\\
~~~$W4$ &  WISE4 mag & $4.79\pm0.03$ & WISE\\
\multicolumn{4}{l}{\hspace{-0.2cm} Library Parameters:}\\
~~~$T_{\mathrm{eff}}$ &  Effective temperature in \unit{K} & 3448 $\pm$ 60 & Yee\\
~~~$\mathrm{[Fe/H]}$ &  Metallicity in dex & -0.02 $\pm$ 0.08 & Yee\\
~~~$\log(g)$ & Surface gravity in cgs units & 4.88 $\pm$ 0.05 & Yee\\
~~~$M_*$ &  Mass in $M_{\odot}$ & 0.35 $\pm$ 0.04 & Yee \\
~~~$R_*$ &  Radius in $R_{\odot}$ & 0.36 $\pm$ 0.01 & Yee \\
~~~Age & Age in Gyr & 4.6 $\pm$ 3.3 & Yee \\
\multicolumn{4}{l}{\hspace{-0.2cm} SED Derived Parameters (Utilized in this Work)} \\
~~~$T_{\mathrm{eff}}$ &  Effective temperature in \unit{K} &  3342 $\pm$ 24  & This Work \\
~~~$\mathrm{[Fe/H]}$ &  Metallicity in dex & 0.07$^{+0.07}_{-0.06}$ & This Work\\
~~~$\log(g)$ & Surface gravity in cgs units & 4.87 $\pm$ 0.05 & This Work \\
~~~$M_*$ &  Mass in $M_{\odot}$ & 0.35$^{+0.05}_{-0.04}$ & This Work \\
~~~$R_*$ &  Radius in $R_{\odot}$ &  0.36 $\pm$ 0.01 & This Work \\
~~~$L_*$ &  Luminosity in $L_{\odot}$ & 0.0155 $\pm$ 0.0004 & This Work\\
~~~Age & Age in Gyr & 6.8$^{+4.6}_{-4.7}$ & This Work \\
\multicolumn{4}{l}{\hspace{-0.2cm} Other Stellar Parameters:}           \\
~~~$v \sin i_*$ &  Rotational velocity in \unit{km/s}  & $<$2 & S20\\
~~~$\Delta RV$ &  Absolute radial velocity in \unit{km/s} & 22.3 $\pm$ 0.2 & Gaia DR3\\
\enddata
\tablenotetext{}{References are: Gliese-Jahreiss \citep{gliese79}, Henry-Draper \citep{nesterov95}, Hipparcos \citep{vanleeuwen07}, Luyten \citep{bakos02}, Stassun \citep{stassun18}, 2MASS \citep{cutri03}, Gaia DR3 \citep{gaia_dr3_21}, Bailer-Jones \citep{bailer-jones21}, Green \citep{green19}, APASS \citep{henden18}, WISE \citep{wright10}}
\end{deluxetable*}

S20 measure a rotation period for GJ 251 using ground based photometry obtained from the T90 telescope \citep{rodriguez10}, the Telescopi Joan Oro \citep[TJO;][]{marti23}, Las Cumbres Observatories \citep[LCO;][]{brown13}, and the Wide Angle Search for Planets survey \citep[WASP;][]{pollacco06}. They obtain a value of 122.1$^{+1.9}_{-2.2}$ days.

\section{Periodogram Analysis}\label{sec:periodogram analysis}

\subsection{Generalized Lomb-Scargle Periodograms}
\label{sec:gls_periodograms}

\begin{figure}
    \centering
    \includegraphics[width=0.5\textwidth]{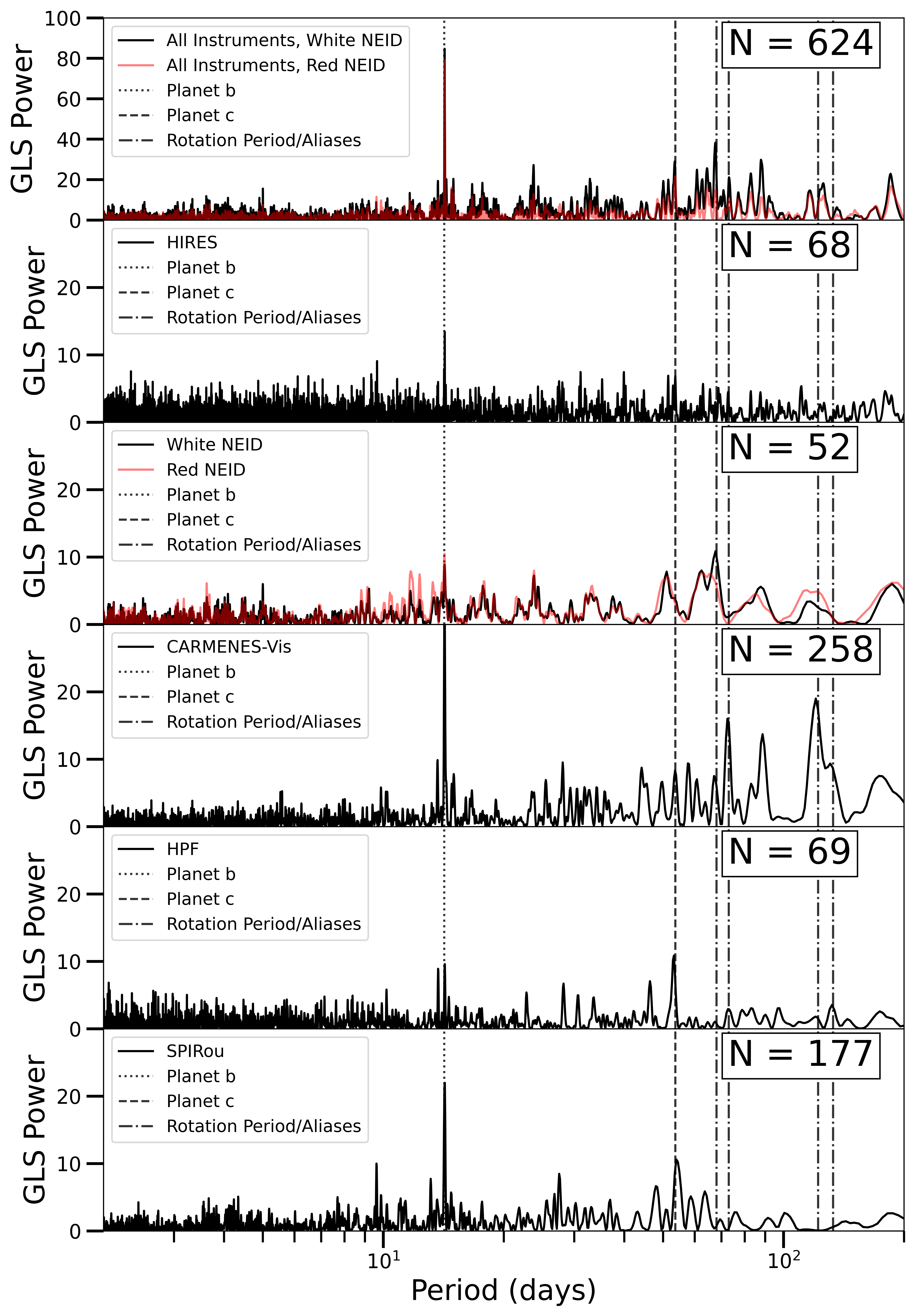}
    \caption{GLS periodograms of GJ 251 RV data. Top: the entire dataset that was used in the analysis of the system, with instrumental offsets applied (taken from Table \ref{tab:2p_posteriors}). Bottom: periodograms on the data from individual instruments. Pre/Post splits for HIRES, HPF, and NEID have been combined after offset correction. Instruments are ordered from bluest to reddest. NEID and all-data periodograms include a red line, indicating the result when red NEID RVs are utilized. Signals of interest highlighted are at 14, 54, 68, 73, 120, and 130 days.}
    \label{fig:gls_by_inst}
\end{figure}

We first analyzed each instrumental dataset separately using a Generalized Lomb-Scargle \citep[GLS;][]{zechmeister09} periodogram. RV data were binned daily before fitting periodograms, and the median value of each instrumental dataset was subtracted before fitting multiple simultaneously.  A plot of the results of a GLS on each instrument is visible in Figure \ref{fig:gls_by_inst}. A signal near 14 days, GJ 251 b, shows up clearly in all datasets. It is the dominant signal in all except for HPF, where its power is barely eclipsed by a signal at 54 days. This 54 day signal has some modest power in all datasets, and appears most strongly in the two infrared instruments, HPF and SPIRou.

\begin{figure}
    \centering
    \includegraphics[width=0.5\textwidth]{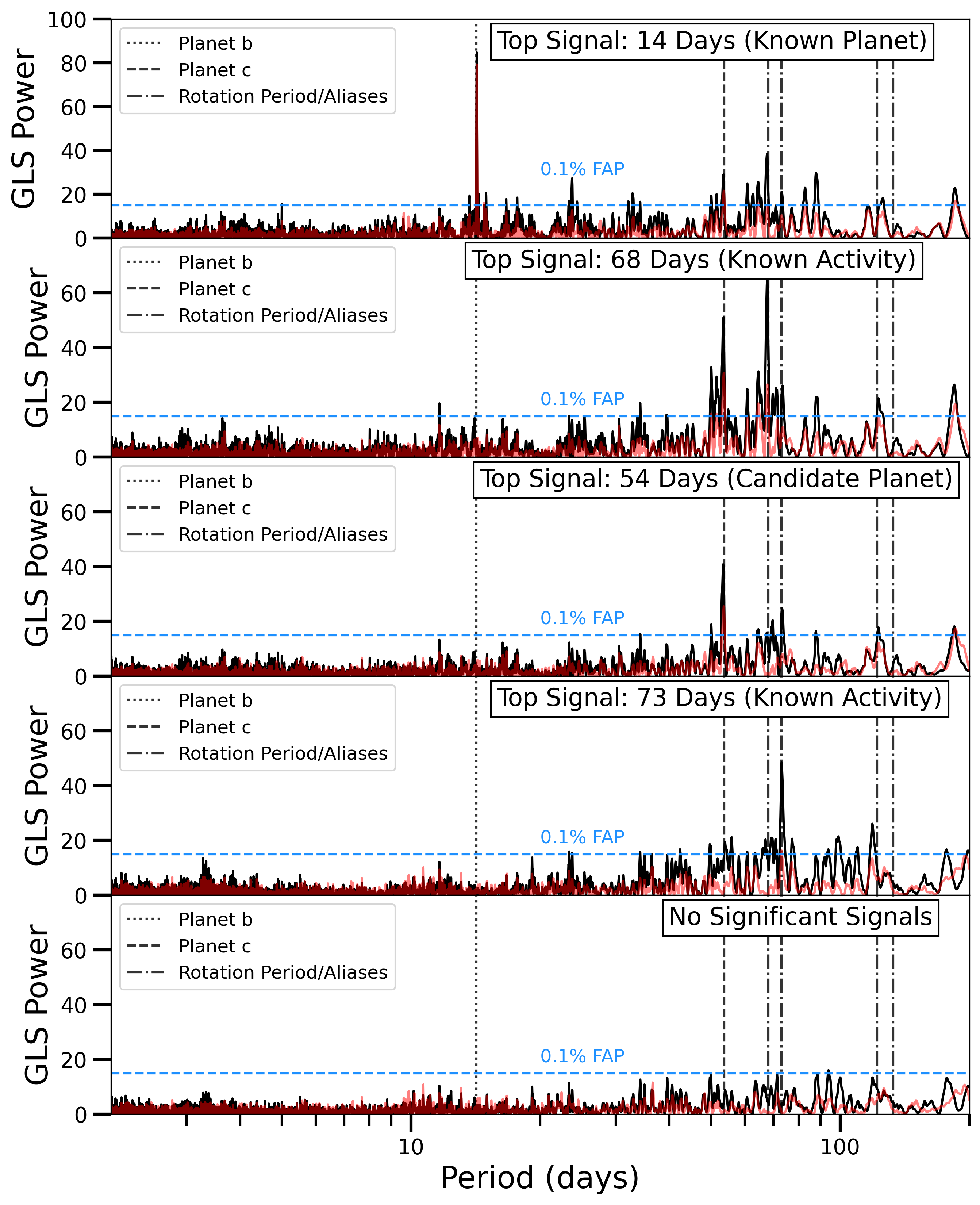}
    \caption{GLS periodograms of GJ 251 RV data as we sequentially remove the most prominent signals. Black plots include white NEID data, while red plots include red NEID data. GJ 251 b is the most significant by far at first, but its subtraction reveals several other prominent signals. Eventually, subtracting signals only adds noise to the periodograms, so we cease our analysis. False alarm probabilities are estimated using methods in \cite{sturrock10}. Signals of interest highlighted are at 14, 54, 68, 73, 120, and 130 days.}
    \label{fig:gls_sig_removal}
\end{figure}

We next perform a series of simple fits to each dominant signal, and subtract this from the RV dataset before performing another GLS analysis (Figure \ref{fig:gls_sig_removal}). After subtracting the dominant 14 day signal, several powerful signals remain. The most prominent is at a period of 68 days, as well as the planet candidate signal at 54 days. Additionally, a few longer-period signals remain prominent. We then subtract the 68 day signal, revealing a dominant signal near 54 days. After the subtraction of the 54 day signal, a 73 day periodicity is dominant. After its subtraction, a forest of signals with similar, modest power appear, with no clearly dominant frequency. This is likely due to noise introduced by the repeated subtraction of simple, imperfect models (sine waves), and we end our GLS analysis.

We additionally perform a periodogram analysis of the RV data window function in Figure \ref{fig:window}, but find no strong peaks other than the lunar period.

\subsection{Agatha Periodograms}

We next utilize the \textsf{Agatha} Bayes Factor periodograms (BFP) outlined in \cite{feng17}. BFPs may offer advantages over traditional GLS periodograms. First, \textsf{Agatha} fits for time-correlated noise using a moving average (MA) model\footnote{Some may not consider this an advantage, as the MA term is a non-physical, ad hoc model, though a simple model comparison in \textsf{Agatha} preferred a single MA term}. Second, \textsf{Agatha} calculates the Bayesian Information Criterion \citep[BIC;][]{kass95}. Though less sophisticated than a complete MCMC or Nested Sampling RV analysis, \textsf{Agatha} provides a first-order model comparison. Third, \textsf{Agatha} fits multiple signals simultaneously, rather than subtracting them iteratively, as in \S \ref{sec:gls_periodograms}. 

\begin{figure}
    \centering
    \includegraphics[width=0.5\textwidth]{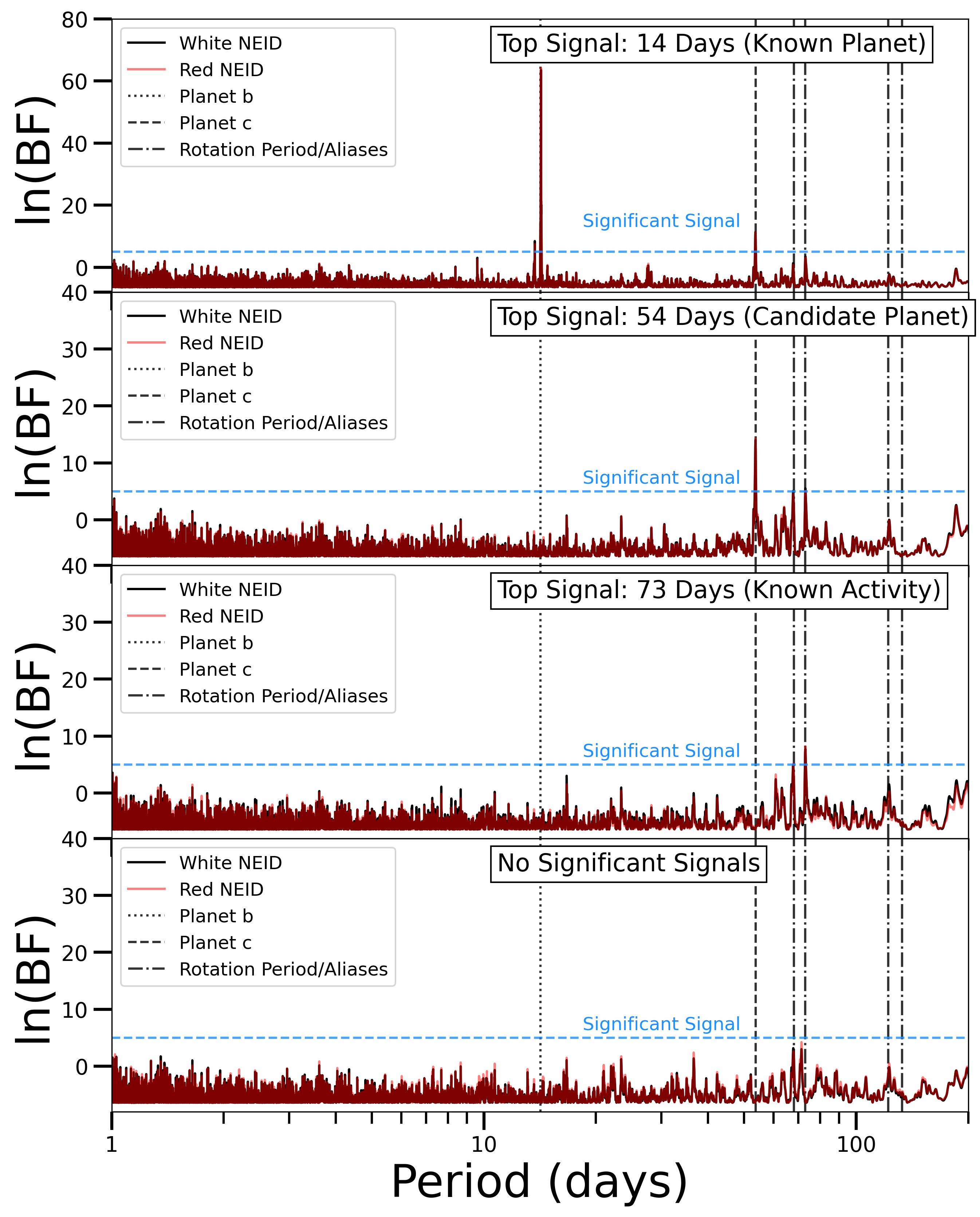}
    \caption{Bayes-Factor Periodograms (BFP) created using the \textsf{Agatha} software package \citep{feng17}. These periodograms use a moving average red-noise model to more accurately identify coherent, periodic signals. The BFP first identifies a very significant signal at 14 days, planet b. It then identifies a significant signal at 53.6 days, our purported planet c. Next, it identifies a 73 day signal which we discuss in \S \ref{sec:interpret}. The final plot identifies no additional significant signals. We designate a Bayes Factor of 5 as the threshold for a significant signal \citep{kass95}. Signals of interest highlighted are at 14, 54, 68, 73, 120, and 130 days.}
    \label{fig:bfp}
\end{figure}

We include a plot of our sequentially fit BFPs in Figure \ref{fig:bfp}. We fit the data using 1 MA component, and an oversampling factor of 10. The 14 day periodicity is clearly identified as the dominant peak, as with the GLS. Departing from the GLS analysis, the 54 day signal is the next most prominent. After fitting the 14 day and 54 day signals, the 73 day signal identified in the GLS periodograms maintains the highest power. When fitting the three top signals simultaneously, a 68 day signal remains. Additional iterations revealed no clearly dominant signals.

Unlike with the GLS analysis, the 14 day and the 54 day signals are the most powerful in the BFP. We theorize that this is due to the inclusion of a simple red-noise model, which may help discriminate against signals that originate from stellar activity. Another distinction is that while the 14 day and 54 day signals both appear with significant power in both GLS and BF periodograms, the 68 day signal never appears strongly in the BFP, suggesting that its source may have been stellar activity. A more detailed analysis is carried out in \S \ref{sec:mcmc}.

\subsection{Stellar Activity Indicator Periodograms}

The relatively low-cadence and uneven temporal sampling of RVs can mask or enhance signals that are not physical. Additionally, the planetary nature of a signal is not always clear, as stellar activity can mimic planetary signals, leading to false positive or false negative detections \citep{robertson14,lubin21}.

\begin{figure*}
    \centering
    \includegraphics[width=\textwidth]{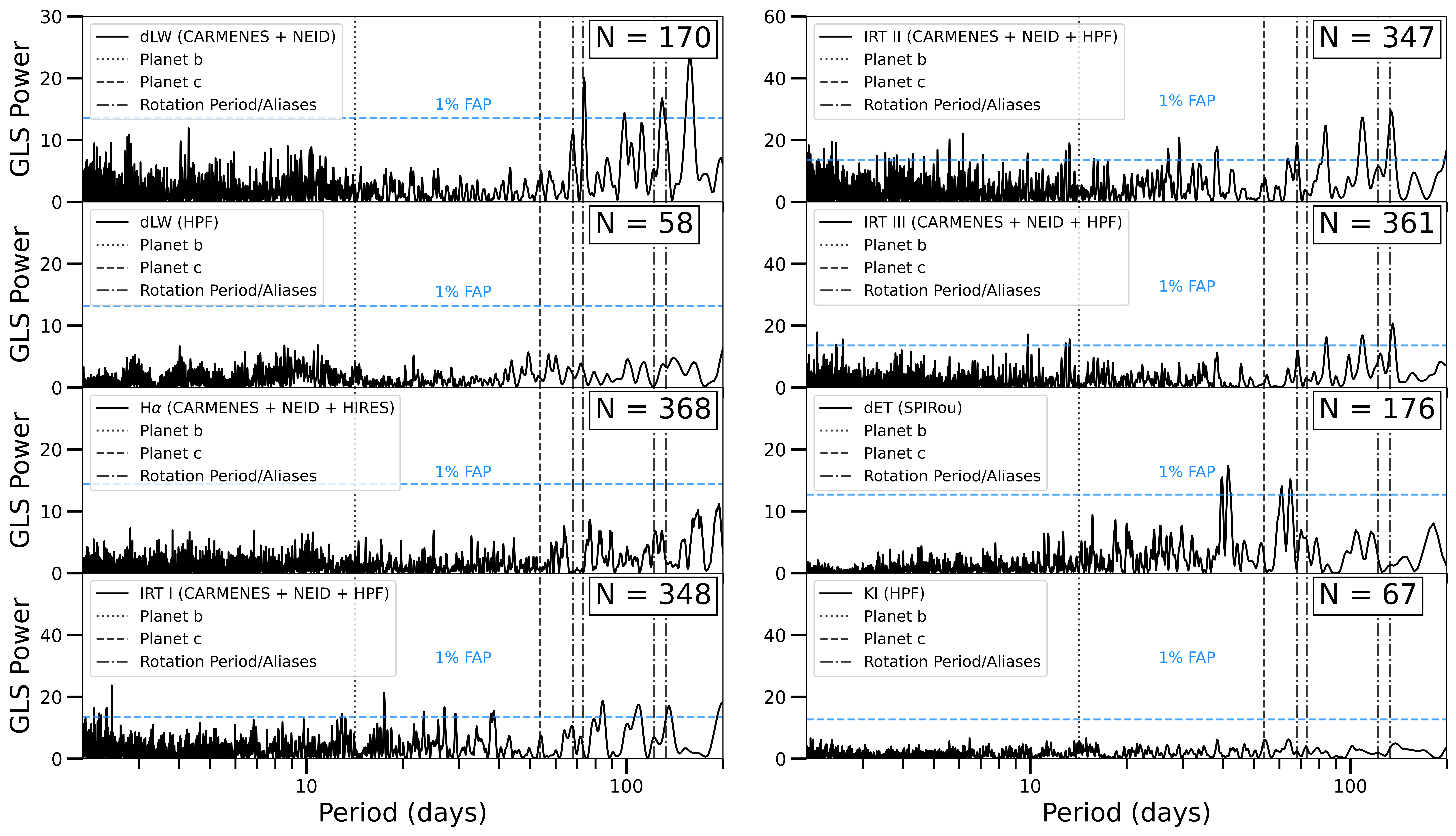}
    \caption{Linewidth, H$\alpha$, Ca IRT, KI, and dET activity indicators of GJ 251. CARMENES/NEID linewidths show strong signals near the rotation period, 68/72 days, and at longer periods. IRT I, II, and III show significant powers near 130 days, likely the rotation period of GJ 251. dET shows significant peaks near 40 and 60 days, likely the second and first harmonics of the rotation period. Significantly, 14 and 54 day signals never appear in the activity indicators, except in KI, though neither are the dominant signal nor significant in power. Signals of interest highlighted are at 14, 54, 68, 73, 120, and 130 days.}
    \label{fig:indicator_periodogram}
\end{figure*}

Stellar activity indicators are one method that can be used to identify a periodic signal as non-planetary in origin. The available GJ 251 data were taken over the course of almost 20 years, and on 5 different instruments. Consequently, there is no universal activity indicator that allows us to combine all instruments. Instead, we choose to analyze several activity indicators, and combine the indicators of different instruments, when possible. We collectively had access to H$\alpha$, Differential Line Widths (dLW), the three lines in the Calcium Infrared Triplet (Ca IRT), the 12435.67 $\rm{\AA}$ potassium line \citep[KI;][Arendtsz et al. in prep]{terrien22}, and a newer indicator, the differential temperature \citep[dET;][]{artigau24}. When making periodograms that use indicators from multiple instruments, we median subtracted each before combining. Due to its redder bandpass, we do not combine dLW from HPF with dLW from NEID/CARMENES. A periodogram analysis of the activity indicators is visible in Figure \ref{fig:indicator_periodogram}.

NEID/CARMENES dLW, IRT I, IRT II, and IRT III indicators consistently show power at 120 or 130 days, likely corresponding to the star's rotation period. The dET indicator shows strong signals at 41.7 days and 64.3 days, likely harmonics of the system's rotation period. A 72 day period shows up strongly in the dLW, and a 68 day period shows up in IRT II, suggesting that these may originate from stellar activity. The 14 day and 54 day signals never appear in the stellar activity periodograms with high power, supporting the hypothesis of a planetary origin for these periodicities.

Throughout our periodogram analysis of RVs and activity indicators, six signals repeatedly appear in the data: a signal at 14 days, 54 days, 68 days, 73 days, 120 days, and 130 days. We discuss these signals at length in \S \ref{sec:interpret}.

\section{MCMC and Nested Sampling Analysis}
\label{sec:mcmc_nested}

\subsection{MCMC Analysis}
\label{sec:mcmc}

We utilized the \texttt{RadVel} software package \citep{fulton18} to fit plantary models to the GJ 251 data. \texttt{RadVel} calculates the orbit of each planet by solving Kepler's equation. Planetary orbits are characterized by 5 free parameters: the velocity semi-amplitude ($K$), the planet orbital period ($P$), the time of periastron ($T_{p}$), the eccentricity ($e$), and the argument of periastron ($\omega$). During our fits, we reparameterize $e$ and $\omega$ as $\sqrt{e}\sin\omega$ and $\sqrt{e}\cos\omega$. This is a frequent choice when modeling eccentricities in order to generate a better behaved posterior \citep{stock20, rosenthal21, macdougall22}. 

The radial velocities of a star include mean and acceleration terms: the mean center of velocity ($\gamma_{i}$), a linear acceleration term ($\dot{\gamma}$), and a second order acceleration term ($\ddot{\gamma}$). The mean center of velocity can be affected by instrumental systematic offsets, and so we adopt an independent $\gamma$ term for each instrument to account for mean RV offsets. We fix the linear and second order acceleration terms ($\dot{\gamma}$ and $\ddot{\gamma}$) to 0. We justify this choice by argument of the length of our timeseries: with over 20 years of data, any significant trend or curvature that might affect our results should be apparent, or too small to detect.

S20 utilized a Gaussian Process \citep[GP;][]{aigrain23} model to account for coherent stellar activity in the RVs during their analysis, though OE23 did not. GPs are often used to model quasiperiodic variability in stars, especially when the amplitude of activity is similar in scale to planetary signals \citep{haywood14,lopezmorales16,cale21}. GPs are stochastic models consisting of a mean function $\mu_{\theta}$ and a covariance matrix, or kernel function, $K_{\alpha}(x_i, x_j)$. $\mu_{\theta}$ represents the physical model of interest, in our case a Keplerian orbit, and is parametrized by the Keplerian parameters outlined above ($\theta$). The covariance matrix, $K_{\alpha}(x_i,x_j)$ models coherent noise, and is parametrized by the hyperparameters of the GP, $\alpha$. The likelihood of a GP model is calculated using Equation \ref{eqn:gp_like}.

\begin{equation}
    \centering
    \label{eqn:gp_like}
    \ln \mathcal{L} = -\frac{1}{2}r_{\theta}^{T}K_{\alpha}^{-1}r_{\theta} - \frac{1}{2} \ln detK_{\alpha} - \frac{N}{2}\ln 2\pi
\end{equation}

Here $r_{\theta}$ is a vector of the residuals of the data minus the model (y - $\mu_{\theta}$). N is the dimensionality of the dataset.

We first adopt the QuasiPeriodic GP kernel for this system, as it is widely used to model coherent stellar noise \citep[Equation \ref{eqn:quasiper};][]{haywood14,lopezmorales16,stock20}. This is the same GP kernel used in S20.

\begin{equation}
    \centering
    \small
    \label{eqn:quasiper}
    K_{\alpha}(x_i, x_j) = \eta_{1}^{2} \times \exp(-\frac{|x_i - x_j|^{2}}{\eta_{2}^{2}} - \frac{\sin^{2}(\frac{\pi|x_i - x_j|}{\eta_{3}})}{2\eta_{4}^{2}})
\end{equation}

$\eta_{1}$ represents the amplitude of variability of the GP, $\eta_{2}$ represents the exponential scale length, $\eta_{3}$ represents the periodic scale length (often the rotation period), and $\eta_{4}$ represents the recurrence timescale of the periodic component.

GPs are flexible tools, but can unnecessarily increase the complexity of the models, especially when coherent noise is not highly prominent, or the number of data points is small. \texttt{Radvel}'s GP likelihood modeling incorporates a different likelihood for each dataset, and this can cause low cadence data in particular to adhere too closely to our model, or become ``overfit" \citep{blunt23,beard24}. While the HIRES dataset is relatively large (68 RVs), it is taken over the course of 16 years, and is very low cadence relative to the other data analyzed here. The HPF and SPIRou data are comparatively high cadence, though they show less variability on timescales associated with stellar activity. Thus, we differ from S20's analysis in this way: we only apply the GP to CARMENES and NEID. Analysis using HIRES, HPF, and SPIRou GP terms tended to suffer from overfitting when attempted. In particular, models would often reduce the instrumental jitter term to a minimum, and use the GP to force every datapoint too cleanly to some local maximum in likelihood-space. This treatment of SPIRou data is consistent with OE23, where the authors found that a GP was not needed.

We additionally explore the application of chromatic GP kernels. \cite{cale21} proposed two expansions of the Quasi-Periodic GP (Equation \ref{eqn:quasiper}) that leverage the different wavelength coverage of different instruments. These kernels, as opposed to the implementation of the QP kernel in \textsf{RadVel}, fit all instruments in a single likelihood. This can help prevent overfitting. The first chromatic GP, K$_{J1}$, introduces an amplitude hyperparameter for each instrument. The form of this kernel is visible in Equation \ref{eqn:KJ1}, where the term in brackets is identical to the exponential portion of the kernel outlined in Equation \ref{eqn:quasiper}.

\begin{equation}
    \centering
    \label{eqn:KJ1}
    K_{J1}(x_i, x_j) = \eta_{s(i)} * \eta_{s^{\prime}(j)} * [...]
\end{equation}

Above, $\eta_{s(i)}$ and $\eta_{s(j)}$ represent the amplitude of the spectrograph s used during the observations taken at times $x_{i}$ and $x_{j}$. The only modification to this kernel is the amplitude term: the covariance between two datapoints is scaled by the product of the GP amplitude associated with the instrument being used at the time of each observation. This is distinct from Equation \ref{eqn:quasiper}, because observations $i$ and $j$ need not come from the same instrument.

\cite{cale21} additionally proposed an alternative GP kernel that enforces the expected scaling between wavelength and RV amplitude when driven by spot contrast. This kernel is outlined in Equation \ref{eqn:KJ2}.

\begin{equation}
    \centering
    \label{eqn:KJ2}
    K_{J2}(x_i, x_j) = \eta_{0}^{2} * \bigg(\frac{\lambda_{0}}{\sqrt{\lambda_{i}\lambda_{j}}}\bigg)^{2\eta_{\lambda}} * [...]
\end{equation}

Again, the exponential portion of this kernel is identical to Equation \ref{eqn:quasiper}, and abbreviated in brackets. The amplitude term introduces two new free parameters, though the number is fixed no matter how many instruments are utilized. $\eta_{0}$ is the GP amplitude in m s$^{-1}$ at the reference wavelength $\lambda_{0}$, which is arbitrary. $\lambda_{i}$ and $\lambda_{j}$ are the associated wavelengths of the spectrograph used for observations $i$ and $j$, respectively. We follow the methods from \S 3.3.2 in \cite{cale21} and take the mean wavelength of an instrument's wavelength coverage to be its characteristic wavelength (weighting equally over the range).\footnote{We caution readers that this is a zeroth order estimate, and a deeper method might be more relevant in future works.} We use their values for HIRES (565 nm), CARMENES-Vis (750 nm), SPIRou (1650 nm), and the reference wavelength (565 nm). We compute values of 1043 nm, 655 nm, and 865 nm for HPF, NEID, and red NEID, respectively. The final hyperparameter, $\eta_{\lambda}$, is an additional power-law scaling parameter to give the GP more flexibility when enforcing a relationship between RV amplitude and instrument wavelength.

We fit RV models of GJ 251 that utilized all three GP kernels. Models were fit that utilized both red and white NEID data. To model the chromatic kernels, we used a modified version of \textsf{RadVel}\footnote{Available to the public: https://github.com/CCBeard/Chromatic-GP}.

Model priors were made to be broad so as to represent our ignorance of the parameters' true values \citep{angus18}. When several orders of magnitude are feasible for a parameter, we use a Jeffreys prior \citep{figueiredo01}, otherwise we use a linear uniform prior. We follow S20 when putting tight uniform priors on the period of planets b and c. We additionally put uniform priors on $K$ from 0.01 m s$^{-1}$ to 10 m s$^{-1}$ to reflect our prior knowledge: with over 20 years of RVs and an RV RMS $\sim$ 3 m s$^{-1}$, it is not feasible that any planet could exist in the system with an RV amplitude $>$ 10 m s$^{-1}$.

Another exception to our broad priors is the periodic hyperparameter of the GP, $\eta_{3}$. This can usually be interpreted as the rotation period of a star, or its alias. Evidence suggests that GJ 251 is an older, slowly rotating star, and S20 identify a rotation period of $\sim$ 120 days. Our RV baseline is sufficiently long to be able to recover such a long period signal, but a slowly rotating star is more likely to have a small amplitude contribution to the RVs from rotation. Indeed our posteriors (Table \ref{tab:2p_posteriors}) suggest that the activity contribution is small, and very much comparable to the relatively small amplitudes of the planets in question. We constrain the $\eta_{3}$ hyperparameter to a prior that is restricted to longer period values, due to our photometry-informed knowledge of the rotation period. We adopt a uniform prior between 100 and 150 days. A complete list of priors used is visible in Table \ref{tab:priors}.

In order to estimate the posterior probability of our models, we utilized a Markov-Chain Monte Carlo (MCMC) sampler to explore the posterior parameter space. \texttt{RadVel} utilizes the MCMC sampler outlined in \cite{foremanmackey13}. We first used the Powell optimization method to provide an initial starting guess for each parameter \citep{powell98}. We then ran 150 independent chains, and assessed convergence using the Gelman-Rubin statistic \citep[G-R;][]{ford06}. During initialization \texttt{RadVel} perturbs the starting position of each chain by a random amount, scaled by the dimension of the parameter. The sampling was terminated when the chains were sufficiently mixed. Chains are considered well-mixed when the G-R statistic for each parameter is $<$ 1.03, the minimum autocorrelation time factor is $\geq$ 75, the max relative change in autocorrelation time $\leq$ 0.01, and there are $\geq$ 1000 independent draws.

\begin{deluxetable*}{llll}
\tablecaption{Priors Used for Bayesian Model Fits \label{tab:priors}}
\tablehead{\colhead{~~~Parameter Name} &
\colhead{Prior} &
\colhead{Units} & \colhead{Description}
}
\startdata
\sidehead{\textbf{Priors for Planet b:}}
~~~P$_{b}$ & $\mathcal{U}^{a}(14.0, 15.0)$ & days & Period \\
~~~T$_{con, b}$ & $\mathcal{U}(2458626.69 - \frac{P_{b}}{2}, 2458626.69 + \frac{P_b}{2})$ & BJD (days) & Time of Inferior Conjunction \\ 
~~~$\sqrt{e} \cos\omega_{b} $ & $\mathcal{U}(-1, 1)$ & ...  & Eccentricity Reparameterization \\
~~~$\sqrt{e} \sin\omega_{b} $ & $\mathcal{U}(-1, 1)$ & ...  & Eccentricity Reparameterization \\
~~~K$_{b}$ & $\mathcal{U}(0.01,10.0)$ & m s$^{-1}$  & Velocity Semi-amplitude \\
\sidehead{\textbf{Priors for Planet c:}}
~~~P$_{c}$ & $\mathcal{U}(50.0, 60.0)$ & days & Period\\
~~~T$_{con, c}$ & $\mathcal{U}(2458929.0 - \frac{P_{c}}{2}, 2458929.0 + \frac{P_c}{2}))$ & BJD (days) & Time of Inferior Conjunction \\ 
~~~$\sqrt{e} \cos\omega_{c} $ & $\mathcal{U}(-1, 1)$ & ...  & Eccentricity Reparameterization \\
~~~$\sqrt{e} \sin\omega_{c} $ & $\mathcal{U}(-1, 1)$ & ...  & Eccentricity Reparameterization \\
~~~K$_{c}$ & $\mathcal{U}(0.01,10.0)$ & m s$^{-1}$  & Velocity Semi-amplitude \\
\sidehead{\textbf{Quasi-Periodic GP Hyperparameters}}
~~~$\eta_{1, CARMENES}$ & $\mathcal{J}^{b}(0.01,100.0)$ & m s$^{-1}$  & GP Amplitude \\
~~~$\eta_{1, NEID-pre}$ & $\mathcal{J}(0.01,100.0)$ & m s$^{-1}$  & GP Amplitude \\
~~~$\eta_{1, NEID-post}$ & $\mathcal{J}(0.01,100.0)$ & m s$^{-1}$  & GP Amplitude \\
~~~$\eta_{2}$ & $\mathcal{J}(1.0,10000.0)$ & days  & Exponential Scale Length \\
~~~$\eta_{3}$ & $\mathcal{U}(100.0,150.0)$ & days  & Periodic Term \\
~~~$\eta_{4}$ & $\mathcal{U}(0.05,0.6)$ & ...  & Periodic Scale Length \\
\sidehead{\textbf{Chromatic GP Hyperparameters (K$_{J1}$)}}
~~~$\eta_{1, HIRES-pre}$ & $\mathcal{J}(0.01,100.0)$ & m s$^{-1}$  & HIRES-pre GP Amplitude \\
~~~$\eta_{1, HIRES-post}$ & $\mathcal{J}(0.01,100.0)$ & m s$^{-1}$  & HIRES-post GP Amplitude \\
~~~$\eta_{1, CARMENES}$ & $\mathcal{J}(0.01,100.0)$ & m s$^{-1}$  & CARMENES GP Amplitude \\
~~~$\eta_{1, HPF-pre}$ & $\mathcal{J}(0.01,100.0)$ & m s$^{-1}$  & HPF-pre GP Amplitude \\
~~~$\eta_{1, HPF-post}$ & $\mathcal{J}(0.01,100.0)$ & m s$^{-1}$  & HPF-post GP Amplitude \\
~~~$\eta_{1, NEID-pre}$ & $\mathcal{J}(0.01,100.0)$ & m s$^{-1}$  & NEID-pre GP Amplitude \\
~~~$\eta_{1, NEID-post}$ & $\mathcal{J}(0.01,100.0)$ & m s$^{-1}$  & NEID-post GP Amplitude \\
~~~$\eta_{2}$ & $\mathcal{J}(1.0,10000.0)$ & days  & Exponential Scale Length \\
~~~$\eta_{3}$ & $\mathcal{U}(100.0,150.0)$ & days  & Periodic Term \\
~~~$\eta_{4}$ & $\mathcal{U}(0.05,0.6)$ & ...  & Periodic Scale Length \\
\sidehead{\textbf{Chromatic GP Hyperparameters (K$_{J2}$)}}
~~~$\eta_{0}$ & $\mathcal{J}(0.01,100.0)$ & m s$^{-1}$  & GP Amplitude \\
~~~$\eta_{\lambda}$ & $\mathcal{U}(0.3,5.0)$ & ...  & Wavelength Dependence Scale \\
~~~$\eta_{2}$ & $\mathcal{J}(1.0,10000.0)$ & days  & Exponential Scale Length \\
~~~$\eta_{3}$ & $\mathcal{U}(100.0,150.0)$ & days  & Periodic Term \\
~~~$\eta_{4}$ & $\mathcal{U}(0.05,0.6)$ & ...  & Periodic Scale Length \\
\sidehead{\textbf{Instrumental Parameters}}
~~~$\gamma_{\rm{HIRES-pre}}$ & $\mathcal{U}(-100,100)$ & m s$^{-1}$ & HIRES offset, pre-upgrade \\
~~~$\gamma_{\rm{HIRES-post}}$ & $\mathcal{U}(-100,100)$ & m s$^{-1}$ & HIRES offset, post-upgrade \\
~~~$\gamma_{\rm{CARMENES}}$ & $\mathcal{U}(-100,100)$ & m s$^{-1}$ & CARMENES offset \\
~~~$\gamma_{\rm{HPF}}$ & $\mathcal{U}(-100,100)$ & m s$^{-1}$ & HPF offset \\
~~~$\gamma_{\rm{NEID}}$ & $\mathcal{U}(-100,100)$ & m s$^{-1}$ & NEID offset \\
~~~$\sigma_{\rm{HIRES-pre}}$ & $\mathcal{U}(0.01,100)$ & m s$^{-1}$  & Instrumental Jitter, HIRES-pre \\
~~~$\sigma_{\rm{HIRES-post}}$ & $\mathcal{U}(0.01,100)$ & m s$^{-1}$  & Instrumental Jitter, HIRES-post \\
~~~$\sigma_{\rm{CARMENES}}$ & $\mathcal{U}(0.01,100)$ & m s$^{-1}$  & Instrumental Jitter, CARMENES \\
~~~$\sigma_{\rm{HPF}}$ & $\mathcal{U}(0.01,100)$ & m s$^{-1}$  & Instrumental Jitter, HPF \\
~~~$\sigma_{\rm{NEID}}$ & $\mathcal{U}(0.01,100)$ & m s$^{-1}$  & Instrumental Jitter, NEID \\
\enddata
\tablenotetext{a}{$\mathcal{U}$ is a uniform prior with $\mathcal{U}$(lower,upper)}
\tablenotetext{b}{$\mathcal{J}$ is a Jeffreys prior with $\mathcal{J}$(minimum, maximum)
}
\end{deluxetable*}

\begin{figure*}
    \centering
    \includegraphics[width=\textwidth]{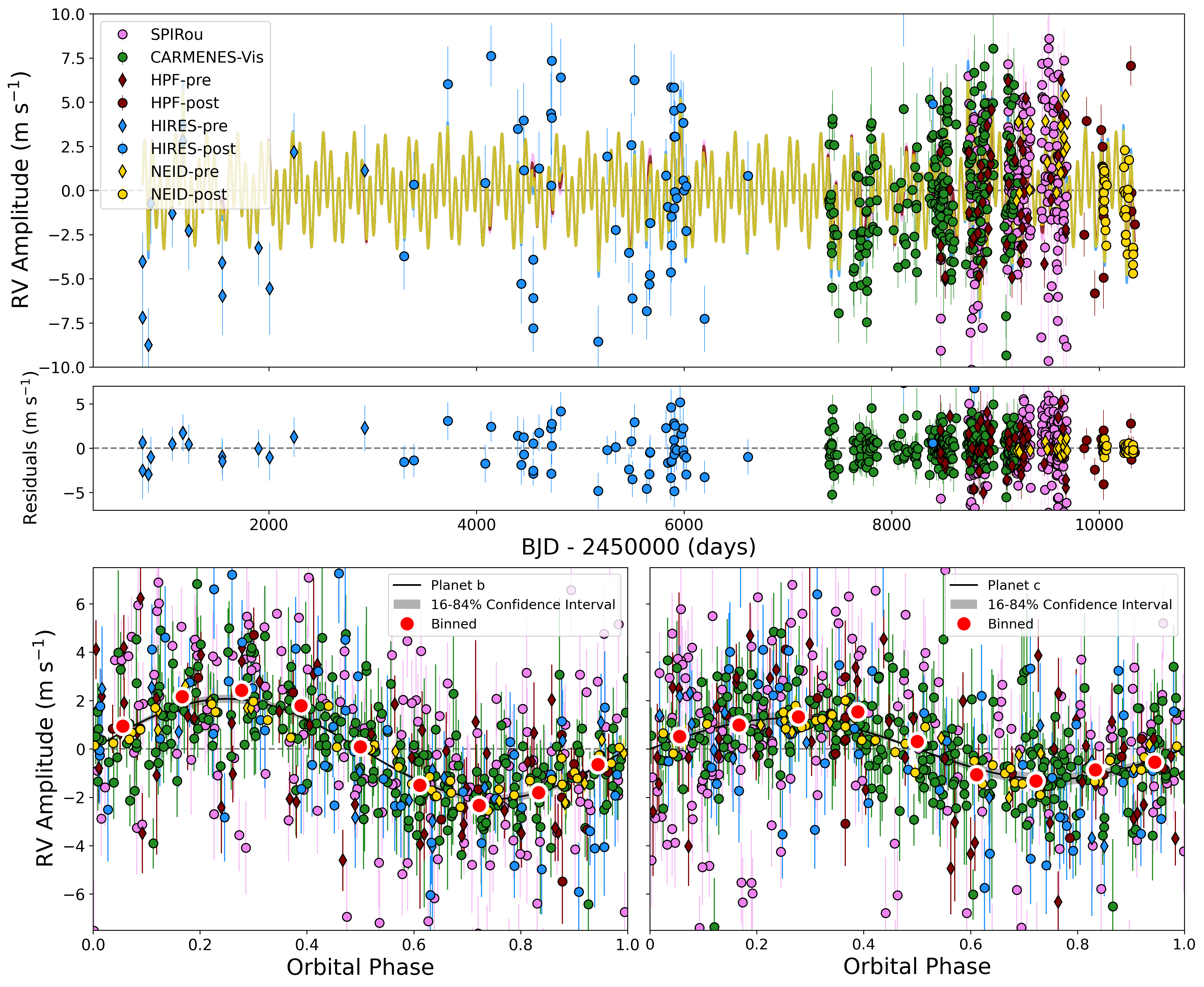}
    \caption{Top: RV timeseries of GJ 251, with GP + 2 planet model overlaid for each instrument. Middle: residuals to a GP + 2 planet model. Bottom: RVs folded to the phase of GJ 251 b and c. RV data are available as data behind the figure.}
    \label{fig:RV_final_white}
\end{figure*}

\begin{figure*}
    \centering
    \includegraphics[width=\textwidth]{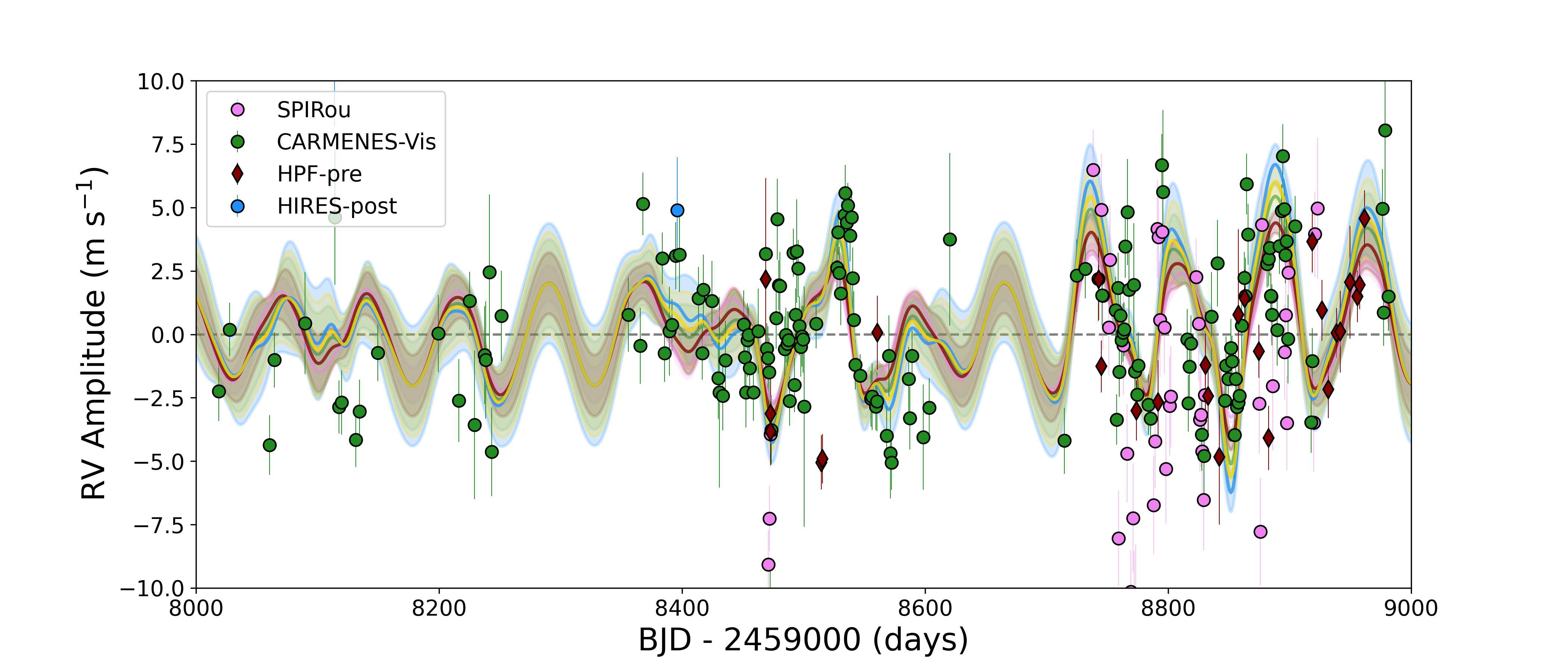}
    \caption{A closer look at the chromatic GP used in our white NEID fits for a particularly dense region of RVs. The solid lines indicate the GP + 2 planet model. Unlike most other GP models, which typically calculate a covariance matrix for each instrument separately, and later combine the likelihoods, the Chromatic kernels for \cite{cale21} calculate covariances across instruments. This relationship is most clear when examining the different predictions from each instrument: they follow the same periodic structure, with deviations mostly due to different amplitudes. Notable also is that the bluer instruments are consistently larger in amplitude, while the redder instruments have smaller GP amplitudes.}
    \label{fig:GPzoom}
\end{figure*}

Plots of our final fits to the GJ 251 RV data are visible in Figures \ref{fig:RV_final_white} and \ref{fig:red_rv_plot}. We additionally include a close zoom on the GP model for a particularly dense region of RV observations in Figure \ref{fig:GPzoom}. 

\normalsize

\normalsize

\subsection{Model Comparison}\label{sec:modelcomparison}

We explored a wide variety of potential models that might explain the data. Consequently, we adopt a formal model comparison to quantitatively estimate which is most likely. We use the Bayesian evidence, sometimes called the marginal likelihood, to choose a preferred model \citep{kass95}. The evidence is a (usually) computationally expensive integral over the likelihood function multiplied by the priors of the model (Equation \ref{eqn:evidence}), and is rarely calculated directly. 

\begin{equation}
    \centering
    \label{eqn:evidence}
    p(D|M_{k}) = \int \mathcal{L}(D|\theta_{k}, M_{k}) * pr(\theta_{k}|M_{k})d\theta_{k}
\end{equation}

Above, $p(D|M_{k})$ is the evidence of model $M_{k}$ given data D, $\mathcal{L}(D|\theta_{k}, M_{k})$ is the likelihood of $M_{k}$ evaluated with parameters $\theta_{k}$ using the data D, and $pr(\theta_{k}|M_{k})$ is the prior probability of the parameters. In general, $\theta_{k}$ is a vector of all free parameters in model $M_{k}$, and thus this can become a very high dimensional integral.

More often the evidence is discussed in relation to the log evidence, which itself is typically a large negative number. Larger (more positive) values of the log evidence are considered more probable, and the difference between two log evidences is defined as the Bayes Factor (BF; Equation \ref{eqn:bayesfactor}), which we can use to quantify the probability of one model over another.

\begin{equation}
    \centering
    \label{eqn:bayesfactor}
    B_{1,2} = \frac{p(D|M_{1})}{p(D|M_{2})}
\end{equation}

A BF of 5 is often considered the minimum value required to prefer a more complex model over a simpler one \citep{kass95}.

To calculate the log evidence we used the \textsf{Juliet} python package \citep{espinoza19}, which exploits the advantages of nested sampling to estimate the log evidence for each model. Nested sampling techniques generate multiple ellipsoids in parameter space that shrink around the best parameter values. The result is a good estimate of the volume in parameter space, replacing our need to take the integral explicitly. The use of multiple ellipsoids allows us to capture even complicated, multimodal distributions, though GJ 251's parameters are generally well behaved (Figure \ref{fig:corner}). We specifically utilized the dynamic nested sampling option, which uses \texttt{Dynesty} to better estimate posterior values \citep{speagle20}. In order to accommodate the chromatic GP kernels outlined in \cite{cale21}, we modified the \textsf{Juliet} source code slightly\footnote{Available to the public: \url{https://github.com/CCBeard/Chromatic-GP}}.

We utilized 300 live points during our analysis. Due to the stochastic nature of nested sampling algorithms, it is often prudent to measure the log evidence several times, and average the results. Our reported log evidence values in Table \ref{tab:modelcomparison} are the median combination of 100 nested sampling runs performed for each model.

%\bigskip

\begin{deluxetable*}{cccccccc}
\tablecaption{Model Comparisons \label{tab:modelcomparison}}
\tablehead{\colhead{Fit}  &  \colhead{No. Parameters}
& \colhead{$\ln (Z)$ (White)} & \colhead{Error of Mean (White)} & \colhead{$\ln (Z)$ (Red)} & \colhead{Error of Mean (Red)} & \colhead{$\Delta \ln (Z)$ (White)} & \colhead{$\Delta \ln (Z)$ (Red)}}
\startdata
0 Planet & 16 & -1723.12 & 0.08 & -1723.61 & 0.08 & -159.95 & -152.19 \\
0 Planet + QP & 20 & -1658.27 & 0.8 & -1666.85 & 1.11 & -95.1 & -95.43 \\
0 Planet + K$_{J1}$ & 27 & -1650.96 & 0.6 & -1657.26 & 1.15 & -87.8 & -85.83 \\
0 Planet + K$_{J2}$ & 21 & -1644.56 & 0.45 & -1650.52 & 0.4 & -81.39 & -79.1 \\
----- & ------ & ----- & ----- & ----- & ----- & ----- & ----- \\
1 Planet & 19 & -1646.9 & 0.2 & -1645.09 & 0.21 & -83.73 & -73.67 \\
1 Planet eccentric (b) & 21 & -1650.07 & 0.27 & -1649.17 & 0.2 & -86.91 & -77.75 \\
1 Planet + QP & 23 & -1575.5 & 1.21 & -1586.53 & 1.24 & -12.33 & -15.11 \\
1 Planet + QP eccentric (b) & 25 & -1581.41 & 3.02 & -1593.98 & 0.6 & -18.25 & -22.56 \\
1 Planet + K$_{J1}$ & 30 & -1575.81 & 1.04 & -1583.53 & 1.13 & -12.64 & -12.11 \\
1 Planet + K$_{J1}$ eccentric (b) & 32 & -1584.31 & 0.82 & -1586.52 & 0.56 & -21.14 & -15.1 \\
1 Planet + K$_{J2}$ & 24 & -1569.32 & 0.39 & -1575.35 & 1.32 & -6.16 & -3.93 \\
1 Planet + K$_{J2}$ eccentric (b) & 26 & -1570.13 & 0.41 & -1582.33 & 0.97 & -6.96 & -10.91 \\
----- & ------ & ----- & ----- & ----- & ----- & ----- & ----- \\
2 Planet & 22 & -1626.23 & 0.27 & -1625.57 & 0.42 & -63.06 & -54.15 \\
2 Planet eccentric (b) & 24 & -1630.3 & 0.31 & -1628.35 & 0.77 & -67.13 & -56.93 \\
2 Planet eccentric (c) & 24 & -1628.69 & 0.52 & -1626.53 & 0.49 & -65.53 & -55.11 \\
2 Planet eccentric (b,c) & 26 & -1633.62 & 2.6 & -1632.01 & 0.65 & -70.45 & -60.58 \\
2 Planet + QP & 26 & -1577.07 & 2.4 & -1586.04 & 1.64 & -13.91 & -14.62 \\
2 Planet + QP eccentric (b) & 28 & -1571.44 & 6.39 & -1596.32 & 10.34 & -8.27 & -24.9 \\
2 Planet + QP eccentric (c) & 28 & -1588.62 & 3.63 & -1597.43 & 4.15 & -25.46 & -26.01 \\
2 Planet + QP eccentric (b,c) & 30 & -1584.47 & 5.7 & -1604.69 & 3.54 & -21.31 & -33.26 \\
2 Planet + K$_{J1}$ & 33 & -1572.5 & 1.94 & -1578.77 & 1.76 & -9.34 & -7.35 \\
2 Planet + K$_{J1}$ eccentric (b) & 35 & -1575.93 & 0.93 & -1584.54 & 1.04 & -12.77 & -13.12 \\
2 Planet + K$_{J1}$ eccentric (c) & 35 & -1580.51 & 1.01 & -1586.02 & 6.72 & -17.35 & -14.6 \\
2 Planet + K$_{J1}$ eccentric (b,c) & 37 & -1579.14 & 6.7 & -1590.51 & 4.68 & -15.98 & -19.09 \\
\textbf{2 Planet + K$_{J2}$} & \textbf{27} & \textbf{-1563.16} & \textbf{0.57} & \textbf{-1571.42} & \textbf{0.62} & \textbf{0.0} & \textbf{0.0} \\
2 Planet + K$_{J2}$ eccentric (c) & 29 & -1567.08 & 0.67 & -1576.34 & 0.73 & -3.92 & -4.92 \\
2 Planet + K$_{J2}$ eccentric (b) & 29 & -1567.08 & 0.67 & -1576.34 & 0.73 & -3.92 & -4.92 \\
2 Planet + K$_{J2}$ eccentric (b,c) & 31 & -1570.46 & 0.64 & -1580.85 & 0.92 & -7.3 & -9.43 \\
\enddata
\end{deluxetable*}

\subsection{Analysis Results}

MCMC fits were generally well behaved. Two planet models were able to constrain the mass of the candidate second planet to better than 5$\sigma$ confidence, and we report improved constraints on the mass of the known planet, GJ 251 b. We find that a two-planet model utilizing the $K_{J2}$ GP kernel is preferred by both the red and white datasets. A summary of the posterior values of our best model for each dataset is provided in Table \ref{tab:2p_posteriors}.

\begin{deluxetable*}{lccl}
\tablecaption{Posteriors of the Preferred Models \label{tab:2p_posteriors}}
\tablehead{\colhead{~~~Parameter} &
\colhead{White NEID} & \colhead{Red NEID} & \colhead{Description}
}
\startdata
\sidehead{\textbf{Orbital Parameters:}}
\sidehead{~~~\textit{Planet b:}}
~~~~~~ $\rm{P_{b}}$ (days) $\dotfill$ & 14.2370 $\pm$ 0.0015 & 14.2382 $\pm$ 0.0015 &   Orbital Period $\dotfill$  \\
~~~~~~$\rm{T_{con,b}}$ (BJD$\textsubscript{TDB}$) $\dotfill$ & 2458626.81 $\pm$ 0.13 & 2458626.89 $\pm$ 0.13 & Time of Inferior Conjunction $\dotfill$ \\
~~~~~~$\rm{e_{b}}$ $\dotfill$ & 0 (fixed) & 0 (fixed) &  Eccentricity $\dotfill$ \\
~~~~~~$\rm{\omega_{b}}$ (degrees) $\dotfill$ & 0 (fixed) & 0 (fixed) &  Argument of Periastron $\dotfill$  \\
~~~~~~$\rm{K_{b}}$ (m s$^{-1}$) $\dotfill$ & 2.01 $\pm$ 0.12 & 2.13 $\pm$ 0.13 &  RV Semi-Amplitude $\dotfill$ \\
\sidehead{~~~\textit{Planet c:}}
~~~~~~$\rm{P_{c}}$ (days) $\dotfill$ & 53.647 $\pm$ 0.044 & 53.65$^{+0.05}_{-0.04}$ &  Orbital Period $\dotfill$ \\
~~~~~~$\rm{T_{con,c}}$ (BJD$\textsubscript{TDB}$) $\dotfill$ & 2458922.0 $\pm$ 1.4 & 2458921.8 $\pm$ 1.4 &  Time of Inferior Conjunction $\dotfill$ \\
~~~~~~$\rm{e_{c}}$ $\dotfill$ & 0 (fixed) & 0 (fixed) &  Eccentricity $\dotfill$ \\
~~~~~~$\rm{\omega_{c}}$ (degrees) $\dotfill$ & 0 (fixed) & 0 (fixed) &  Argument of Periastron $\dotfill$ \\
~~~~~~$\rm{K_{c}}$ (m s$^{-1}$) $\dotfill$ & 1.23$\pm$0.22 & 1.27 $\pm$ 0.21  & RV Semi-Amplitude $\dotfill$ \\
\sidehead{\textbf{Planetary Parameters:}}
\sidehead{~~~\textit{Planet b:}}
~~~~~~$\rm{M \sin i _{b}}$ (M$_\oplus$) $\dotfill$ & 3.85$^{+0.35}_{-0.33}$ & 4.07 $\pm$ 0.37 & Minimum Mass $\dotfill$ \\
~~~~~~$\rm{a_{b}}$ (AU) $\dotfill$ & 0.0808 $\pm$ 0.0042 & 0.0808 $\pm$ 0.0042  & Semi-major Axis$\dotfill$\\
~~~~~~$\rm{\langle F_{b} \rangle}$ ($\unit{W/m^2}$)$\dotfill$ & 3225 $\pm$ 351 & 3225 $\pm$ 347 & Average Incident Flux$\dotfill$ \\
~~~~~~$\rm{S_{b}}$ (S$_\oplus$)$\dotfill$ &  2.37 $\pm$ 0.26 & 2.37 $\pm$ 0.25 & Planetary Insolation$\dotfill$ \\
~~~~~~$\rm{T_{eq,b}}$ (K)$\dotfill$ & 336 $\pm$ 1 & 290 $\pm$ 8 & Equilibrium Temperature$^{a}$ $\dotfill$ \\
\sidehead{~~~\textit{Planet c:}}
~~~~~~$\rm{M \sin i _{c}}$ (M$_\oplus$) $\dotfill$ & 3.88 $\pm$ 0.79 & 3.80 $\pm$ 0.73 & Minimum Mass $\dotfill$ \\
~~~~~~$\rm{a_{c}}$ (AU) $\dotfill$ & 0.196 $\pm$ 0.014 & 0.196 $\pm$ 0.022 & Semi-major Axis$\dotfill$ \\
~~~~~~$\rm{\langle F_{c} \rangle}$ ($\unit{W/m^2}$)$\dotfill$ &  550 $\pm$ 81  & 550 $\pm$ 124 & Average Incident Flux$\dotfill$   \\
~~~~~~$\rm{S_{c}}$ (S$_\oplus$)$\dotfill$ & 0.404 $\pm$ 0.059 & 0.404 $\pm$ 0.091 & Planetary Insolation$\dotfill$ \\
~~~~~~$\rm{T_{eq,c}}$ (K)$\dotfill$ & 216 $\pm$ 0.8 & 186 $\pm$ 10 & Equilibrium Temperature$^{a}$ $\dotfill$  \\
\sidehead{\textbf{GP Hyperparameters}}
~~~$\eta_{0}$ (m s$^{-1}$) $\dotfill$ & 2.57$^{+0.64}_{-0.45}$ & 2.49$^{+0.66}_{-0.46}$ & GP Amplitude ($K_{J2}$) $\dotfill$ \\
~~~$\eta_{\lambda}$ $\dotfill$ & 1.23$^{+0.92}_{-0.61}$ & 1.27$^{+0.90}_{-0.64}$ & Wavelength Dependence Scale $\dotfill$ \\
~~~$\eta_{2}$ (days) $\dotfill$ & 89$^{+56}_{-75}$ & 95$^{+59}_{-81}$  & Exponential Scale Length $\dotfill$\\
~~~$\eta_{3}$ (days) $\dotfill$ & 143$^{+22}_{-10}$ & 141$^{+18}_{-8}$ & Periodic Term $\dotfill$ \\
~~~$\eta_{4}$ $\dotfill$ & 0.25$^{+0.11}_{-0.05}$ & 0.24$^{+0.10}_{-0.05}$ & Periodic Scale Length $\dotfill$ \\
\sidehead{\textbf{Instrumental Parameters}}
\sidehead{~~~\textit{RV Jitter}}
~~~~~~$\rm{\sigma_{HIRES-pre}}$ (m s$^{-1}$)  $\dotfill$ & 1.02$^{+1.32}_{-0.71}$ & 0.97$^{+1.2}_{-0.66}$ & HIRES-pre Jitter \\
~~~~~~$\rm{\sigma_{HIRES-post}}$ (m s$^{-1}$)  $\dotfill$ & 2.48$^{+0.66}_{-0.67}$ & 2.55$^{+0.66}_{-0.69}$ & HIRES-post Jitter \\
~~~~~~$\rm{\sigma_{CARMENES-Vis}}$ (m s$^{-1}$)  $\dotfill$ & 0.91 $\pm$ 0.15 & 0.92 $\pm$ 0.15 & CARMENES-Vis Jitter \\
~~~~~~$\rm{\sigma_{SPIRou}}$ (m s$^{-1}$)  $\dotfill$ & 3.39$^{+0.26}_{-0.25}$ & 3.41$^{+0.26}_{-0.25}$ & SPIRou Jitter \\
~~~~~~$\rm{\sigma_{HPF-pre}}$ (m s$^{-1}$) $\dotfill$ & 1.98$^{+0.33}_{-0.30}$ & 1.99$^{+0.33}_{-0.30}$ & HPF-pre Jitter \\
~~~~~~$\rm{\sigma_{HPF-post}}$ (m s$^{-1}$) $\dotfill$ & 1.32$^{+0.73}_{-0.66}$ & 1.25$^{+0.73}_{-0.65}$ & HPF-post Jitter \\
~~~~~~$\rm{\sigma_{NEID-pre}}$ (m s$^{-1}$)  $\dotfill$ & 1.03$^{+0.50}_{-0.37}$ & 1.31$^{+0.51}_{-0.41}$ & NEID-pre Jitter \\
~~~~~~$\rm{\sigma_{NEID-post}}$ (m s$^{-1}$)  $\dotfill$ & 0.46$^{+0.12}_{-0.10}$  & 0.64$^{+0.18}_{-0.17}$ & NEID-post Jitter \\
\sidehead{~~~\textit{RV Offset}}
~~~~~~$\rm{\gamma_{HIRES-pre}}$ (m s$^{-1}$)  $\dotfill$ & -1.61 $\pm$ 1.13 & -1.7 $\pm$ 1.1 & HIRES-pre Offset \\
~~~~~~$\rm{\gamma_{HIRES-post}}$ (m s$^{-1}$)  $\dotfill$ & 0.80$^{+0.71}_{-0.69}$ & 0.78 $\pm$ 0.69 & HIRES-post Offset \\
~~~~~~$\rm{\gamma_{CARMENES-Vis}}$ (m s$^{-1}$)  $\dotfill$ & -0.14$^{+0.33}_{-0.34}$ &  -0.14$^{+0.32}_{-0.33}$ & CARMENES-Vis Offset \\
~~~~~~$\rm{\gamma_{SPIRou}}$ (m s$^{-1}$)  $\dotfill$ & -0.28$^{+0.36}_{-0.40}$ & -0.26$^{+0.36}_{-0.40}$ & SPIRou Offset \\
~~~~~~$\rm{\gamma_{HPF-pre}}$ (m s$^{-1}$)  $\dotfill$ & -0.36$^{+0.38}_{-0.40}$ & -0.31$^{0.38}_{0.40}$ & HPF-pre Offset \\
~~~~~~$\rm{\gamma_{HPF-post}}$ (m s$^{-1}$)  $\dotfill$ & -0.02$^{+0.59}_{-0.62}$ & 0.06 $\pm$ 0.58 & HPF-post Offset \\
~~~~~~$\rm{\gamma_{NEID-pre}}$ (m s$^{-1}$)  $\dotfill$ & 1.67$^{+0.67}_{-0.71}$ & 1.68$^{+0.65}_{-0.68}$& NEID-pre Offset \\
~~~~~~$\rm{\gamma_{NEID-post}}$ (m s$^{-1}$)  $\dotfill$ & -0.05 $\pm$ 0.69 & 0.15 $\pm$ 0.16 & NEID-post Offset  \\
\enddata
\tablenotetext{a}{Estimated assuming an albedo of 0.1}
\normalsize
\end{deluxetable*}

\section{Interpreting the Signals in GJ 251}
\label{sec:interpret}

In our analysis, we highlight six signals of interest, each outlined in detail below. These signals appear repeatedly in periodogram analysis, and all warrant investigation. 

\subsection{14 Day Signal}
\label{sec:14}
S20 and OE23 both identified the 14 day signal in their analyses, and concluded that its origin was planetary. Our analysis is in agreement with both. The 14 day periodicity is by far the strongest signal in the RV data, and no strong periodicities corresponding to 14 days appear in activity indicators.

All RV datasets, except for HPF, identify the 14 day signal as the most prominent (Figure \ref{fig:gls_by_inst}). HPF still detects the signal strongly, and this is probably explained by the 1 day alias near the 14 day signal that is visible in all of the datasets, discussed in S20. We suspect that, due to the lower quantity of HPF and NEID data, the two are more susceptible to aliases confusing the dominant signal. This results in a smaller power associated with the true planetary signal than would be expected.

We conclude that the 14 day periodicity is indeed a planet, GJ 251 b.

\subsection{122 and 133 Day Signals}
\label{sec:122133}
 S20 identify a stellar rotation period of $\sim$122 days for GJ 251 from photometry. Such a periodicity is present in the RV dataset, seen most strongly in the CARMENES RVs. A 133 day signal appears repeatedly in the stellar activity indicator periodograms, especially IRT II and III. Further, the peak detected in the dLW data is consistent with either a 122 day or 133 day signal. Finally, the dET indicators do not show a strong signal at either period, though 41.7 and 64.3 day periodicities appear strongly, and are the second and first harmonics of $\sim$125 day and $\sim$128 day periods.

A number of analyses suggest that the rotation period of GJ 251 is somewhere between 120 and 130 days. Our failure to more precisely constrain the rotation period is likely a result of the quasiperiodic nature of stellar activity signals, the uneven temporal cadence of RV measurements, and possibly differential rotation.

\subsection{68 and 73 Day Signals}
\label{sec:6873}
A 68 day signal is the strongest signal in the GLS periodograms after the removal of planet b, and a 73 day signal is strongest signal detected by the BFP when simultaneously fitting planet b and the 54 day signal. Subtracting the 73 day signal from the BFP significantly reduces the power of the 68 day signal (Figure \ref{fig:bfp}), suggesting that the two signals may be related.

Both signals appear to inherit most of their power from the CARMENES or NEID datasets (Figure \ref{fig:gls_by_inst}), which seem to be most affected by activity. Furthermore, neither signal is detected with any strength in the HPF or SPIRou data, which appear to be the least susceptible to activity contamination. This suggests that the signals may be related to the stellar rotation period.

The system rotation period is likely near 122 or 133 days. The latter rotation period would place the first harmonic of the rotation period (66.5 days) close to the 68 day signal. The 68 day signal appears with some power in the IRT II and IRT III indicators, though its is not the highest in either. The 73 day signal appears with high power in the CARMENES + NEID dLW indicator plot (Figure \ref{fig:indicator_periodogram}), suggesting that it may be related to stellar activity.

We conclude that the 68 day signal is probably the first harmonic of the stellar rotation period, and that the 73 day signal likely originates from stellar activity.

\subsection{54 Day Signal}
\label{sec:54}

\subsubsection{Meaning of the 54 Day Signal}
A strong signal at 53.6 days is identifiable in the GLS and BF periodograms (Figures \ref{fig:gls_by_inst}, \ref{fig:bfp}). Notable is that this signal seems to originate with the greatest power in the HPF and SPIRou data. Our GLS analysis ranks it behind the 68 day signal as third most significant, though our BFP analysis ranks this signal as second most significant.

One explanation is that the 54 day signal is a small ($\sim$1 m s$^{-1}$ amplitude) planetary signal that was missed in S20 due to activity contamination in the bluer CARMENES visible RVs. Starspots seem to be the primary contaminating feature in the RV timeseries, as the rotation period identified from photometry (and perhaps its aliases/harmonics) shows up frequently during RV analysis. 54 days is not particularly close to the first harmonic associated with either possible rotation period (61 days; 66.6 days), nor is it close to either second harmonic (41 days; 44 days), supporting an origin not related to the stellar rotation period.

HPF and SPIRou's redder bandpasses make them less susceptible to contamination from starspots \citep{crockett12,robertson20}. The fact that the 54 day signal appears most strongly in the IR datasets, and the rotation signals appear most strongly in the visible datasets, is suggestive of a planetary origin for the 54 day signal. At least, the 54 day signal appears more strongly in the red RVs--this might be explained by some stellar activity cycle other than spots that is more prominent in the infrared, or it may be that the small amplitude 54 day planet is recovered more strongly with less contaminating activity signals. This is further supported by the BFP analysis, which identifies the 54 day signal more strongly than the 68, 73, or 122 day signals. This is a sensible result when moving from a GLS (which naively highlights any periodic signal) to a BFP with a MA (which accounts for correlated noise using a red-noise model).

Stellar activity can masquerade as planetary signals, and we cannot eliminate this possibility entirely for the 54 day signal. It is additionally possible that the prominence of the 54 day signal in HPF and SPIRou is not related to their wavelength regime, but due to fortunate sampling occurring in the two infrared datasets. We perform a series of tests in the following sections to more strictly scrutinize the 54 day signal.

\subsubsection{Stability of the 54 Day Signal}
\label{sec:stability}

One key difference between planetary signals and stellar variability is that planetary signals are stable over long periods of time, while stellar variability will often evolve on timescales of a few rotation periods \citep{giles17,gilbertson20}. Our dataset contains a huge quantity ($>$ 600) of RVs that extend over twenty years, allowing us to scrutinize the stability of the 54 day signal over this time.

\begin{figure*}
    \includegraphics[width=\textwidth]{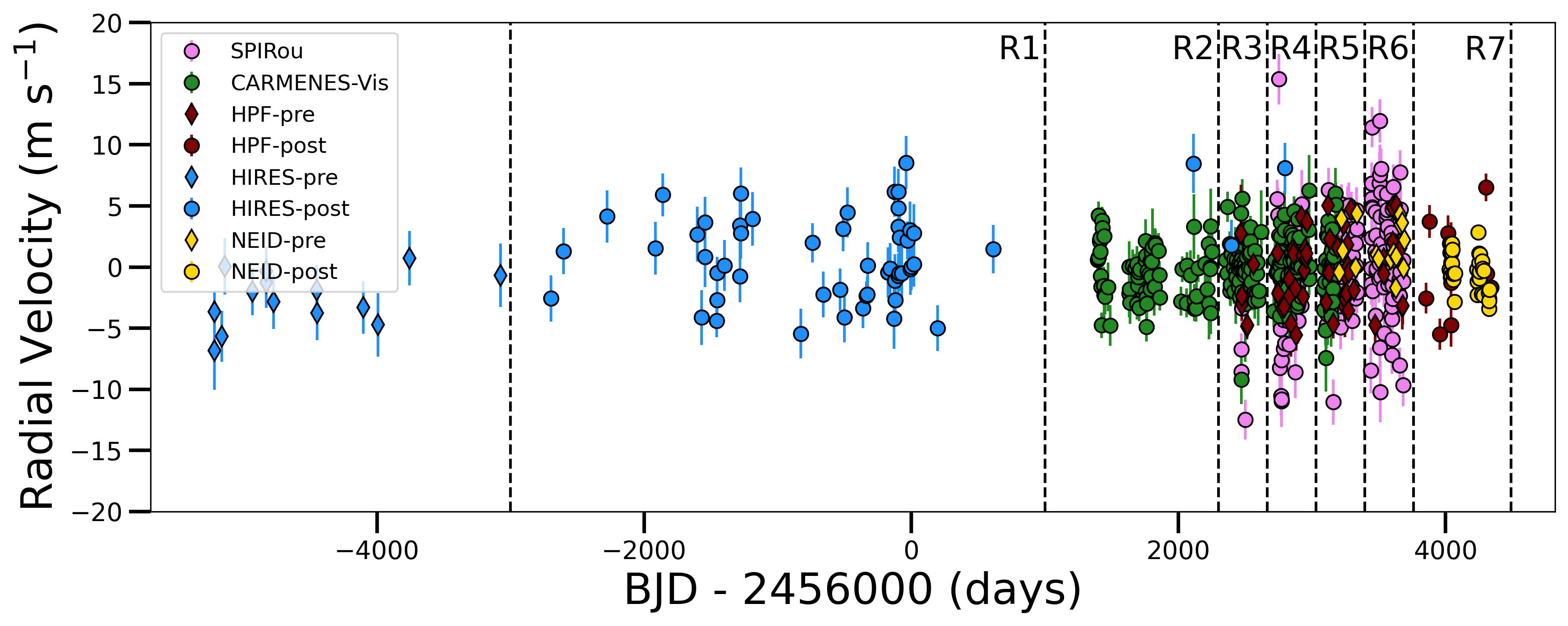}
    \caption{Time series of RVs for GJ 251, showing how we divided observing ``regions" to test the consistency of the candidate planet parameters. We mostly try to divide into observing seasons (R3, R4, R5, R6), though R1 and R7 encompass multiple years of observing due to lower cadence.}
    \label{fig:season_div}
\end{figure*}

We perform two-planet MCMC fits, similar to those in \S \ref{sec:mcmc}, but to subsets of the dataset. First, we perform such fits on each instrument by itself, to see if the signal is greatly changed in any instrument. We next divide the data into seven temporal regions. When the RV data are dense, we pick regions coinciding with observing seasons, each corresponding to a year. Older RV data are of lower cadence, however, and we chose to combine a few observing seasons for older HIRES and CARMENES data. The division of our seasons is visible in Figure \ref{fig:season_div}.

\begin{figure*}
    \includegraphics[width=\textwidth]{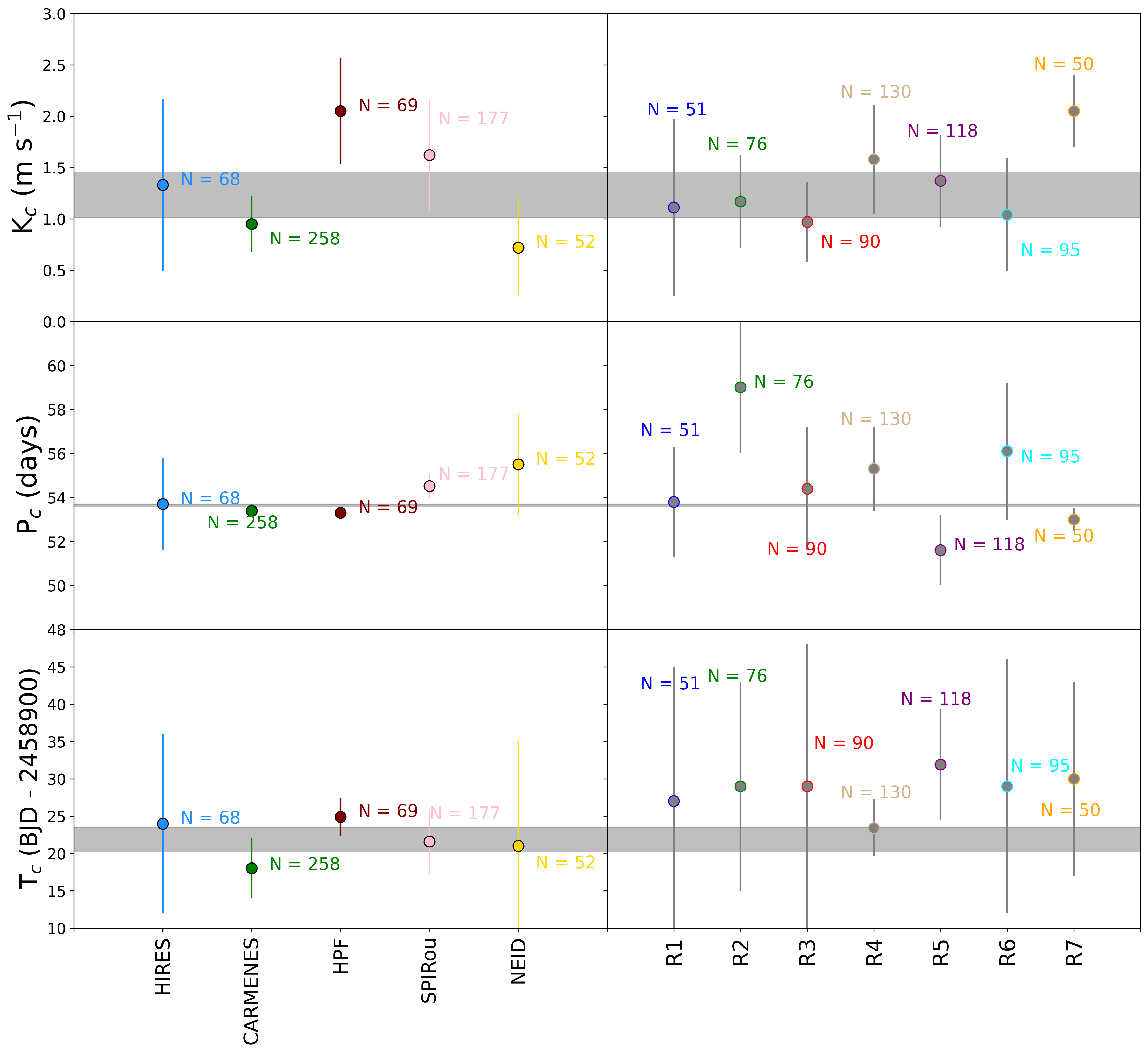}
    \caption{The posterior values for $K_{c}$, $P_{c}$, and $T_{c}$, and their uncertainties, are shown from top to bottom. Fits divided by instrument on the left, and divided by observing region on the right. A gray bar highlights the posterior value of our best all-data fit and its 1$\sigma$ confidence interval.}
    \label{fig:season_results}
\end{figure*}

We fix the instrumental offsets and jitter terms to those found in our preferred fit posteriors (Table \ref{tab:2p_posteriors}), but otherwise left the models free to fit for each planetary and GP parameter. We are most interested in the recovered values of P$_{c}$, K$_{c}$, and T$_{c}$. If the 54 day signal is planetary, these values should be constant over time. Our results are visible in Figure \ref{fig:season_results}.

The instrumental signals generally agree with our fit posteriors, and each other. HIRES and CARMENES data always agree well. The HPF fit is peculiar, in that it recovers a larger amplitude signal at a slightly shorter period. Nonetheless, the values are all 2$\sigma$ consistent with our fit posteriors, and 1$\sigma$ consistent with each other. NEID and SPIRou fits are a bit discrepant in period, but with correspondingly large uncertainties. 

The region fits are even more in agreement. The fits are fully consistent with each other, though the orbital period recoveries show some variance. This is not surprising considering that the period of the candidate planet is sufficiently long to challenge some of the shorter-baseline datasets. Some combination of poor phase coverage and stellar activity very conceivably interferes with our ability to accurately measure the orbital period over only one or two seasons.

Both instrument and region fits are in two sigma or better agreement across all measured parameters for the candidate planet. If the signal were indeed stellar variability, originating from a particular region of time, we might instead expect one or two regions with highly discrepant amplitudes, periods, or phases. The lack of such a detection suggests that the origin of the signal may be planetary.

\subsubsection{False Alarm Probability}
\label{sec:false_alarm}

The known planet, in combination with a quasi-periodic stellar activity signal, might create some alias resembling a second planet. Such a false alarm might be exacerbated by our uneven temporal sampling and the variety of instruments employed in the analysis. 

To test how commonly these quirks of our data recreate a 54 day signal, we generate a series of synthetic datasets. First, we utilize \textsf{RadVel} to generate the RV signal of one and two planet models at each of our observation times using our best model median posterior values. We next generate stellar activity signals at each RV observation time by drawing GP posterior values from the posterior distributions of the top performing activity model, the $\mathcal{K}_{J2}$ GP kernel, and adding these to our one and two planet models. Next, we add white noise by drawing our final synthetic values from a normal distribution centered on each model, with a standard deviation equal to the quadrature sum of the measured RV error at that timestamp, and the posterior jitter estimate for its instrument. Finally, we add instrumental offsets taken from our best model posteriors to create synthetic datasets that closely resemble real observational datasets.

\begin{figure}
    \centering    \includegraphics[width=0.5\textwidth]{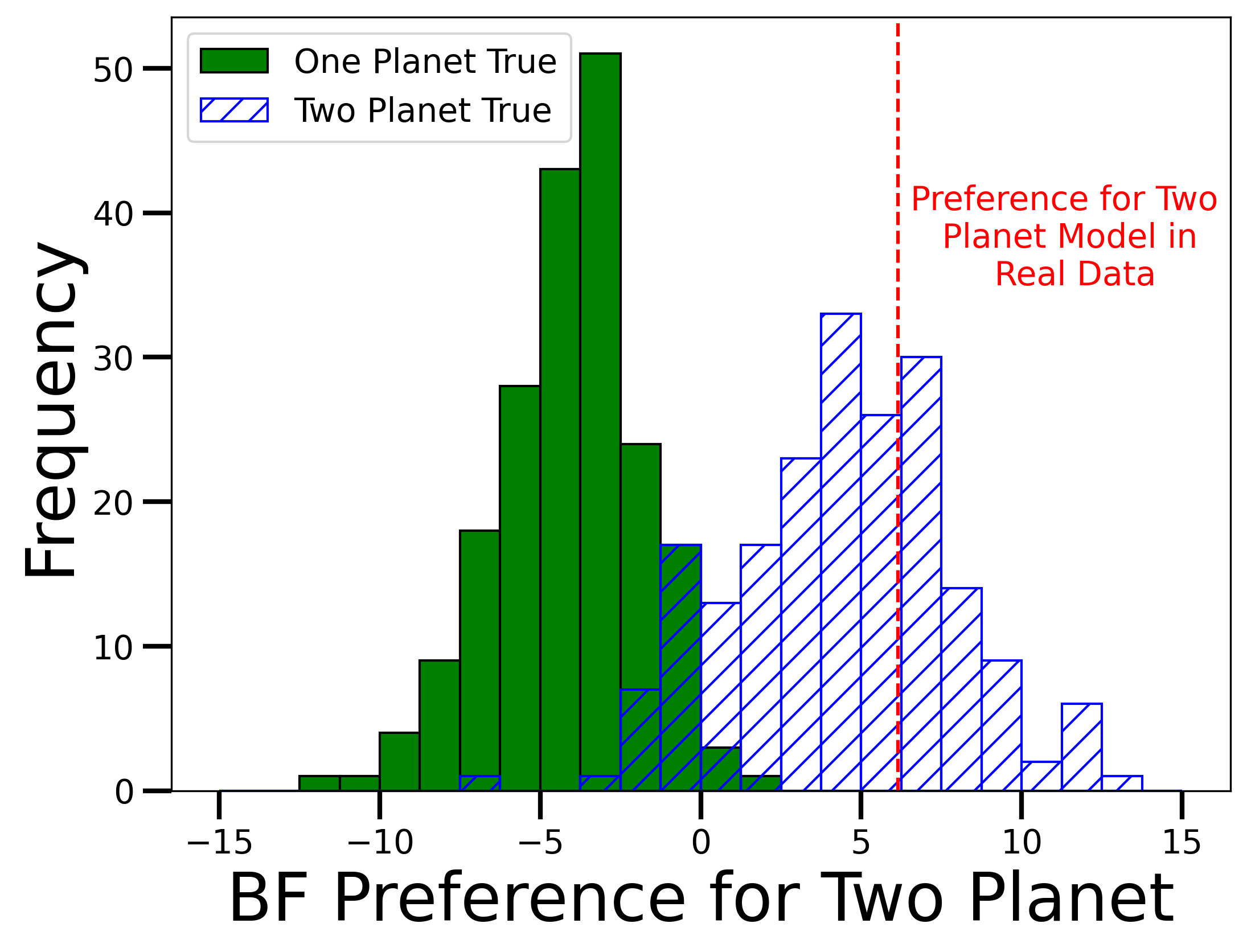}
    \caption{Two histograms highlighting the results of model comparisons on simulated datasets. After generating 200 fake one-planet, and 200 fake two-planet datasets, we performed nested sampling fits for one- and two-planet models, and computed the difference in Bayes Factor. When the fake data include only one injected planet, it is very rare that a two-planet model is preferred. On the other hand, when the generated data include two injected planets, a two-planet model is often preferred, with the BF in the real data falling on the center of this distribution.}
    \label{fig:fap}
\end{figure}

We generated two hundred fake datasets per model, and we performed nested sampling fits with \textsf{Juliet} on each. Once we estimated the evidence, we calculated the BF preference for a two planet model for each synthetic dataset. As in \S \ref{sec:modelcomparison}, we ran these evidence calculations 100 times and took the median. We include a plot of the distribution of BFs in Figure \ref{fig:fap}.

None of our synthetic datasets generated from a one planet model developed a strong preference for two planets. Alternatively, when using a dataset with two planets, a two planet model was often strongly preferred. The level of preference for a two planet model found in the real data is highly anomalous if originating from a one planet + activity model, and is much more consistent with the two planet synthetic datasets. This suggests that the true data more likely represent a two planet model than a one planet model.

We must qualify this result, however. Firstly, we were limited by computational resources in exploring as many run as would have been ideal, utilizing only 200 false datasets. Running such models for even one such dataset takes a great deal of resources, as each requires that we run 100 1-planet and 100 2-planet models to calculate a robust value for the evidence. Additionally, our results are dependent on our choice of activity model, in this case the $\mathcal{K}_{J2}$ GP kernel. We chose this activity model because it was the most preferred during model comparison. It may be that the true activity model is more likely to conspire with the 14 day signal to create a 54 day false positive. This test could be repeated for various GP kernels, though none would be definitive.

\section{Non-RV Datasets}
\label{sec:other_data}

\subsection{Photometry}
\label{sec:transits}
We utilize existing photometric datasets to determine if any planets in GJ 251 are transiting.

\begin{figure}
    \centering
    \includegraphics[width=0.5\textwidth]{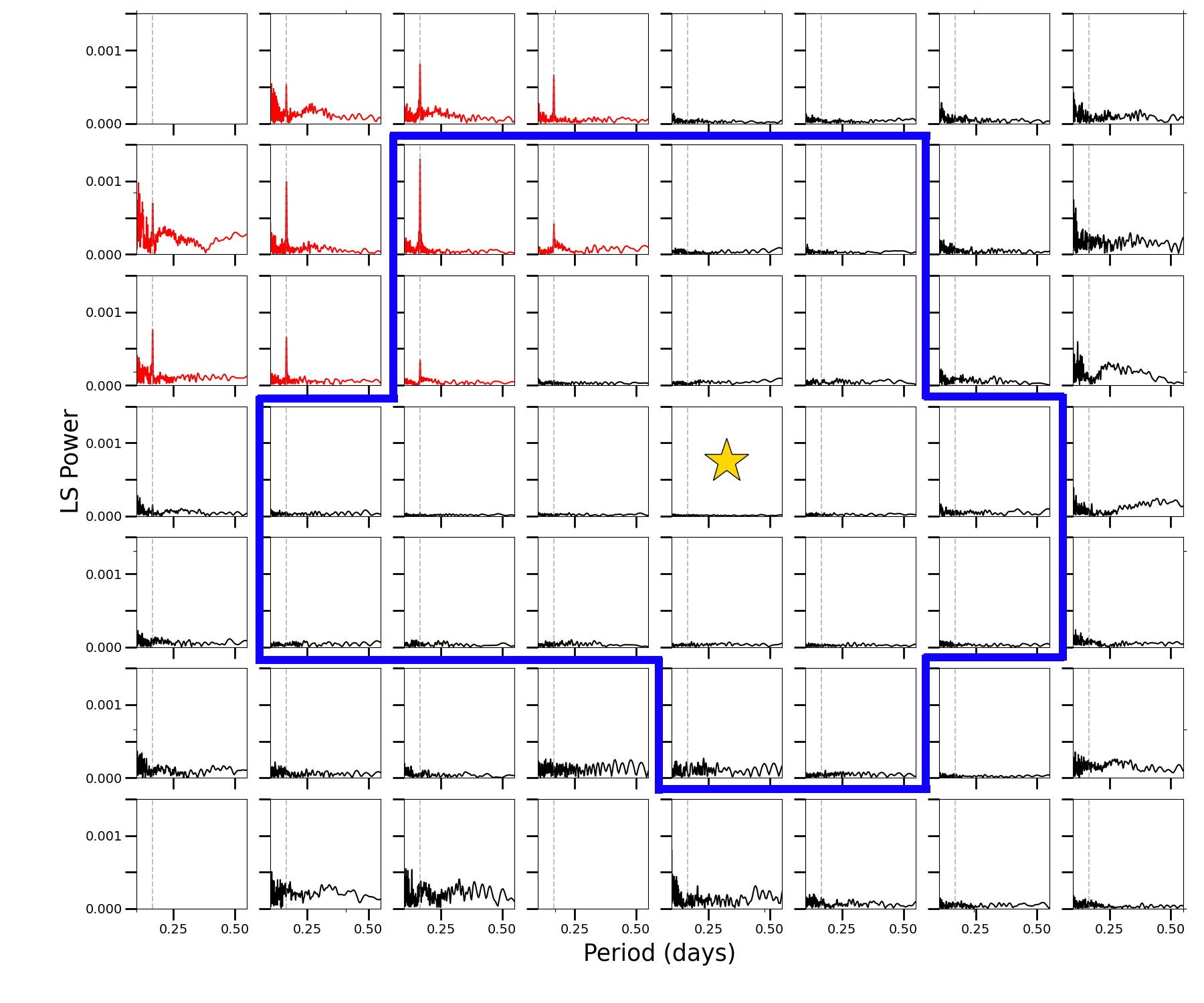}
    \caption{Pixel-by-pixel periodogram of the TESS photometry of GJ 251 during TESS Sector 20. Each plot corresponds to a pixel of the TESS CCD, and the solid blue line is the aperture used by the PDCSAP pipeline during the generation of the GJ 251 lightcurve. A gold star highlights the pixel that contains GJ 251. This plot can help us discover the source of the periodic variability seen in the GJ 251 lightcurve. It is clear that the $\sim$ 4 hour periodicity originates not from GJ 251, but other, neighboring pixels, which we highlight in red. Consequently, we can rule out the signal as fast rotation of GJ 251.}
    \label{fig:PBP_plot}
\end{figure}

GJ 251 was observed by TESS \citep{ricker15} during Sector 20 (2019 December 24 - 2020 January 21), Sector 45 (2021 November 6 - 2021 December 2), Sector 47 (2021 December 30 - 2022 January 28), Sector 60 (2022 December 23 - 2023 January 18), Sector 71 (2023 October 16 - 2023 November 11), and Sector 72 (2023 November 11 - 2023 December 7). Examining the pre-search data conditioned simple aperture flux \citep[PDCSAP;][]{caldwell20} reveals peculiar short-timescale periodicity on the order of 4 hours, which might suggest rapid stellar rotation. This would seriously conflict with our previous determination of GJ 251 as an older, slowly rotating star in \S \ref{sec:periodogram analysis}. We created a pixel-by-pixel plot of periodograms of GJ 251's TESS photometry in Figure \ref{fig:PBP_plot}. Periodogram analysis of each pixel reveals that this signal is likely caused by a nearby star periodically contaminating the aperture, and not by any variability intrinsic to GJ 251.

\begin{figure*}
    \centering
    
    \includegraphics[width=\textwidth]{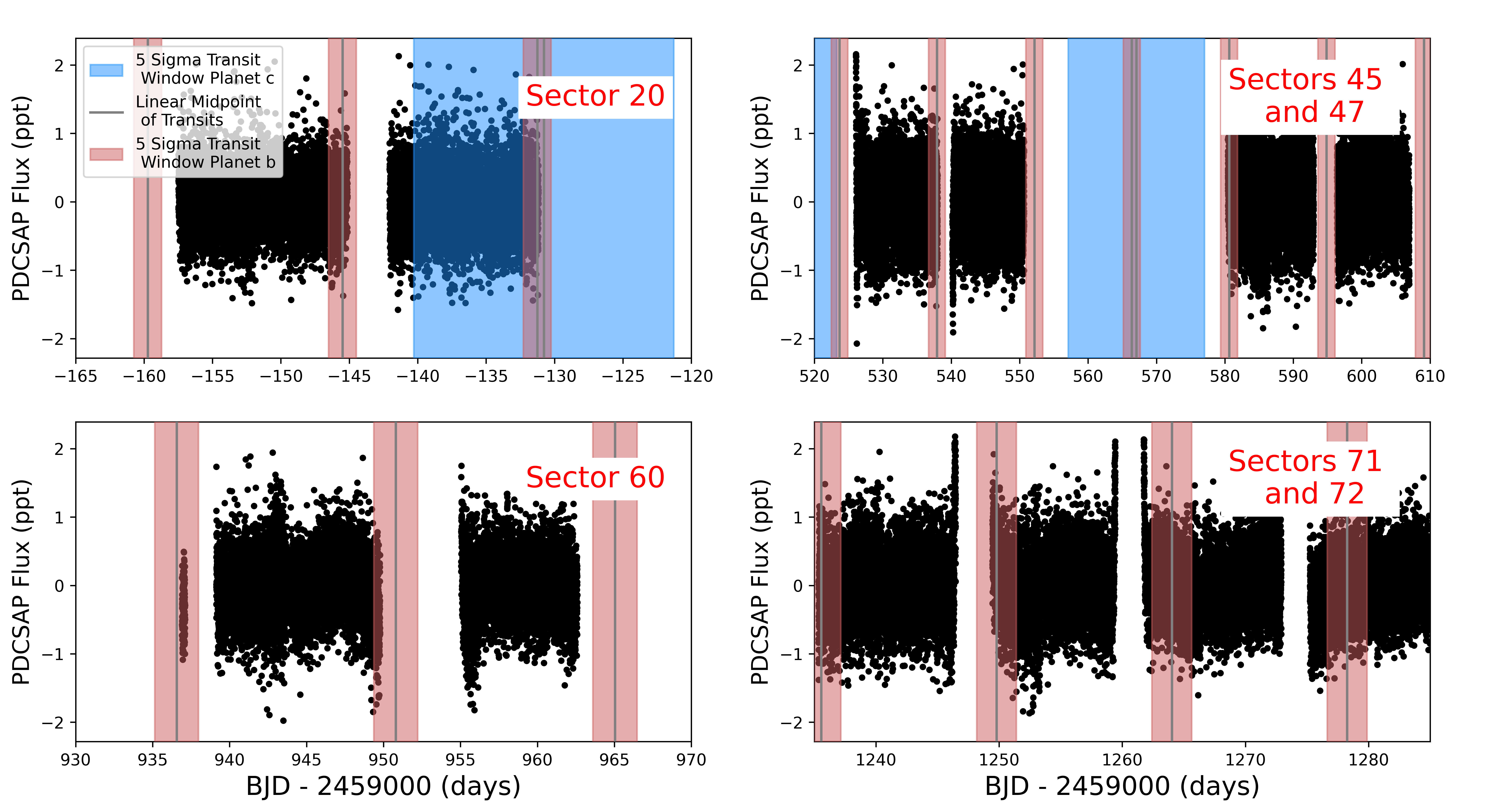}
    \caption{Short cadence photometry of GJ 251 taken by TESS during sectors 20, 45, 47, 60, 71 and 72. Overlaid is the transit ephemeris of GJ 251 b and c as determined from our RV fits. Sectors 71 and 72 strongly suggest that GJ 251 b does not transit. GJ 251 c, with its 54 day period and less precise ephemeris, is difficult to constrain with short-baseline TESS photometry. One predicted transit time does fall into Sector 20 photometry, though the uncertainty is so large as to be fairly uninformative. We detect no transits in the TESS photometry of GJ 251.}
    \label{fig:transit_times}
\end{figure*}

Using the ephemeris obtained from our best fit model, we predict transit times for GJ 251 b and c. We include a plot of the six available TESS Sectors' photometry and our transit predictions. A combination of short Sector lengths and the unfortunate coincidence that planet b's period is nearly half of a TESS Sector, while the candidate planet's period is nearly double a TESS sector, frequently placed expected transit times in data gaps. We fit a box least squares \citep[BLS;][]{kovacs02} transit search algorithm on TESS PDCSAP flux across all Sectors, but no significant transits were identified. Figure \ref{fig:transit_times} shows that a few transits of GJ 251 b should have appeared in the photometry, if the planet indeed transits, though the predicted transit times of GJ 251 c typically fall in data gaps.

\subsection{Astrometry}
\label{sec:astrometric}

It is possible that either planetary signal might be explained by a highly inclined massive object. One way to investigate this possibility is to look for astrometric motion of the host star in the plane of the sky. Gaia DR3 \citep{gaia_dr3_21} measures the renormalized unit weight error \citep[RUWE; ][]{lindegren21} to estimate the goodness of fit to the general astrometric solution. A RUWE of 1.0 is considered a perfect fit for a single star, and it is suggested that a value $<$ 2.0 (or 1.4 by some stricter metrics) should be used as a cutoff for significant astrometric motion. GJ 251 has a measured RUWE of 1.15, which is consistent with no detected astrometric motion. We additionally compared Hipparco-Gaia astrometric measurement differences and found that the solution in \cite{brandt18} is fully consistent with no unseen astrometric motion.

\section{Discussion}\label{sec:discussion}

\subsection{A Habitable Zone Super-Earth Amenable to Direct Imaging in Reflected Light}
\label{sec:direct_imaging}

\begin{figure}
    \centering
    \includegraphics[width=0.5\textwidth]{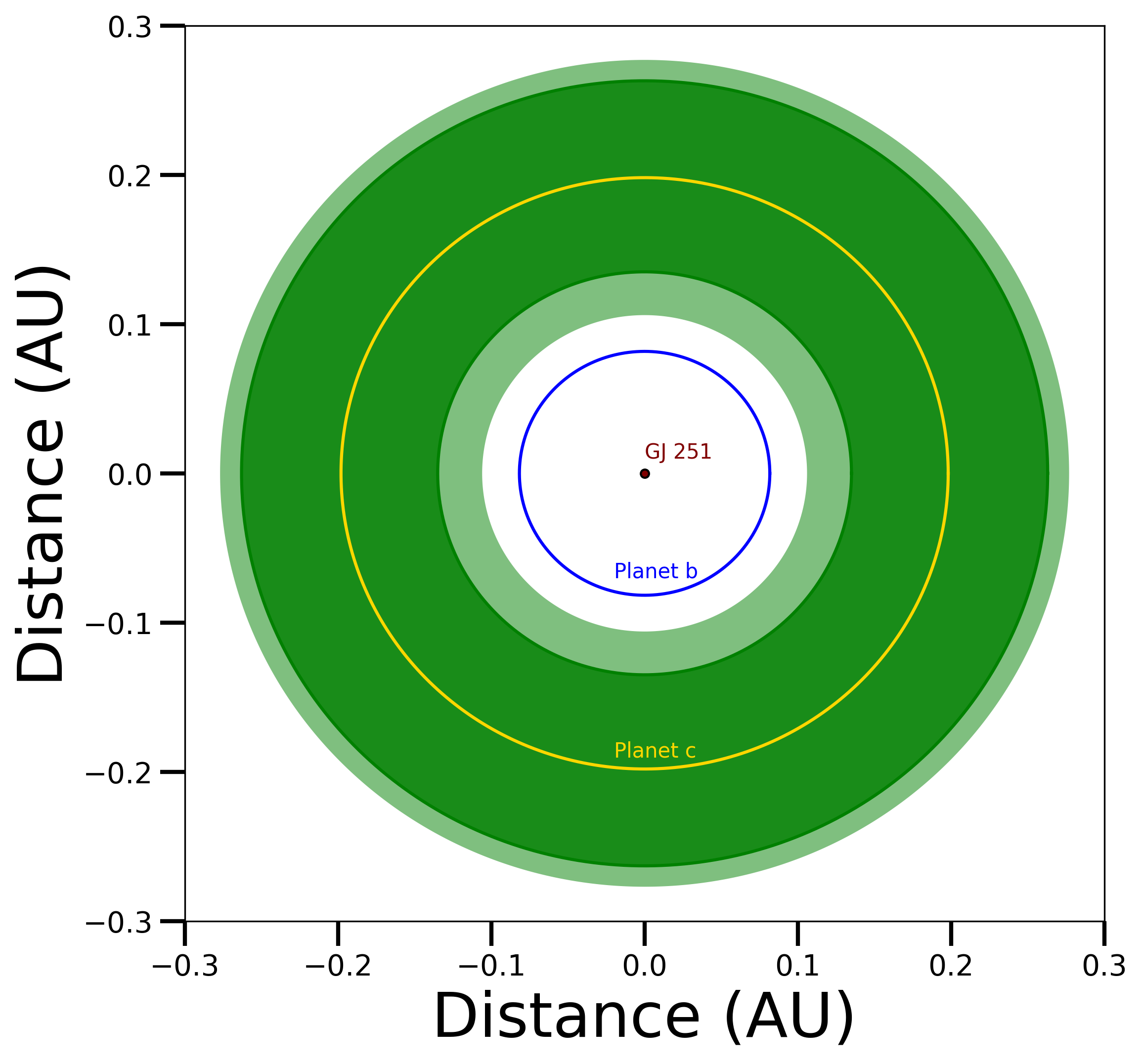}
    \caption{The size of GJ 251 and the orbits of GJ 251 b and c. Stellar size and orbital distances are to scale. In light green we show the optimistic HZ, and in dark green the conservative HZ \citep{kopparapu13,kopparapu14}. GJ 251 c orbits directly in a regime where liquid water might be able to exist on its surface.}
    \label{fig:HZ}
\end{figure}

GJ 251 c would be a prime place to search for life. GJ 251 c orbits in the conservative Habitable Zone \citep[HZ;][]{kopparapu13,kopparapu14,kane16} of its host star (Figure \ref{fig:HZ}), and its measured minimum mass is consistent with a terrestrial planet (we adopt a threshold $m\sin i$ $<$ 5.0 M$_{\oplus}$). Furthermore, the proximity of GJ 251 to Earth makes GJ 251 c perhaps the top target for imaging a HZ planet in the celestial Northern Hemisphere.

GJ 251 is the 74$^{th}$ closest star to Earth \citep{bailer-jones21}, and the 22$^{nd}$ closest known planet-hosting star (NASA Exoplanet Archive). Of these 22, only 8 have positive declinations (Barnard's Star \citep{basant25}; Ross 128 \citep{bonfils18}; GL 725A \citep{corteszuleta25}; GJ 15A \citep{pinamonti18}; Teegarden's Star \citep{zechmeister19}; GJ 9066 \citep{quirrenbach22}; GJ 687 \citep{burt14,rosenthal21}), and only one (GJ 15A) has a higher stellar effective temperature than GJ 251.

\begin{figure*}
    \centering
    \includegraphics[width=\textwidth]{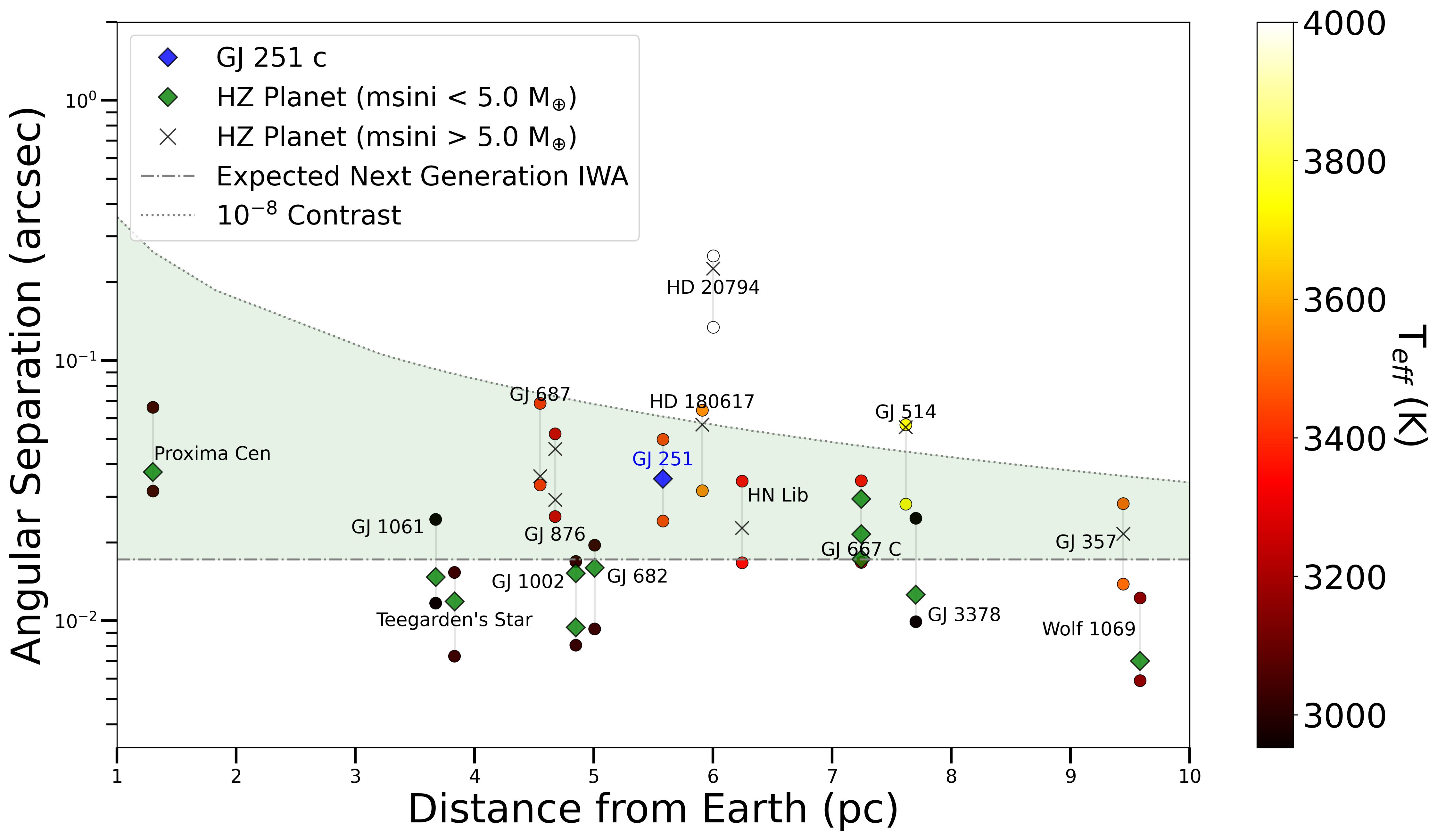}
    \caption{Angular separation of the HZ for the closest stars to Earth that have known HZ planets. The conservative inner and outer edges of the system's HZ are plotted and connected by a gray line. Points are color-coded by stellar effective temperature. Planets that might be terrestrial are highlighted with green diamonds, while gas-dominant planets are marked with an x. We separate these populations at a minimum mass of 5 $M_{\oplus}$, as planets with greater mass become increasingly unlikely to remain terrestrial. We further include a plausible end-goal IWA for next generation thirty meter class telescopes and a curve showing the angle at which a 2 R$_{\oplus}$ planet would be visible at 10$^{-8}$ contrast with an albedo of 0.5. This curve actually corresponds to a particular physical separation, which correspondingly translates into smaller angular separations for systems further from Earth.} The region where both conditions are met is highlighted in green.
    \label{fig:direct_imaging}
\end{figure*}

A higher stellar effective temperature can alleviate some of the challenges when directly imaging a planet. In order to directly image an exoplanet via coronagraphy, the planet-to-star contrast ratio must exceed the coronagraph raw contrast, and a planet must have a projected separation between the coronagraph's inner working angle (IWA) and outer working angle \citep[OWA;][]{guimond18}. A serious challenge arises when imaging the HZ of stars, as the HZ is much closer than typically imaged, long period exoplanets \citep{marois08, macintosh15}. Additionally, many of the planets of interest are more mature planets that require observations using reflected light, rather than emitted. Even dedicated extreme adaptive optics systems on 30 m class telescopes will only be able to access HZs of the closest stars to Earth. GJ 251 is one such star, among a few, highlighted in Figure \ref{fig:direct_imaging}\footnote{Data taken from the NASA Exoplanet Archive on 02/15/2025 (DOI: https://doi.org/10.26133/NEA1); contrast upper limits depend on several assumptions including planet radius and albedo and should not be viewed as a hard upper limit on image-ability.}.

\normalsize

At its maximum estimated orbital distance, GJ 251 c would have a semi-major axis of 0.196$\pm$0.014 AU assuming minimum mass. Because of its proximity to Earth, the planet maintains an angular separation of 0.035$^{\prime \prime}$. In fact, as seen in Figure \ref{fig:direct_imaging}, the entire HZ of GJ 251 is theoretically imagable with next generation telescopes, opening the possibility to identification of other planets in the system's HZ.

GJ 251 is not a young star (6.8$^{+4.6}_{-4.7}$ Gyr), and any terrestrial planets in the system are not likely to have maintained heat of formation. Infrared imaging of GJ 251 c via emitted light might be possible in the N band if the temperature exceeds $\sim$ 250 K \citep{quanz15}, though reflected light imaging is more feasible. We estimate the contrast of GJ 251 c using Equation \ref{eqn:contrast} \citep{li21}.

\begin{equation}
    \label{eqn:contrast}
    \centering
    \epsilon = A * \frac{1}{\pi} * \frac{R_{p}^2}{a^{2}}
\end{equation}

A is the albedo of the planet, R$_{p}$ is the planet radius, and a the semi-major axis. Here we assume that the planet could be imaged at its maximal sky-projected orbital separation, and we have removed wavelength dependence of albedo for simplicity, as a wavelength dependent investigation is beyond the scope of this paper. We estimate a contrast for the planet candidate using the non-parametric mass-radius relationship from \textsf{MRExo} \citep{kanodia19}, finding an estimated radius of 1.81 R$_{\oplus}$. Such a radius would likely correspond to a sub-Neptune composition. Consulting \cite{parc24}, a planet with an Earth-like composition would have a radius $\sim$1.5 R$_{\oplus}$. Combined with our calculated orbital separation (Table \ref{tab:2p_posteriors}), we estimate, assuming an albedo of 0.5, contrast in the 1.81 R$_{\oplus}$ case of $2.4 \times 10^{-8}$, and a contrast in the 1.5 R$_{\oplus}$ case of $1.7 \times 10^{-8}$. More pessimistic radii and albedo values (1.1 R$_{\oplus}$, albedo = 0.1) might produce contrasts as low as $1.8 \times 10^{-9}$.

With an angular separation of 0.035$^{\prime \prime}$ and contrasts as low as $1.8 \times 10^{-9}$, GJ 251 c is currently impossible to directly image. Future observatories are likely to change this.

The Planetary Sciences Imager \citep[PSI; ][]{guyon18} is a next generation instrument being considered for the Thirty Meter Telescope \citep[TMT; ][]{skidmore15,konopacky23}, an ELT that may see first light in the next ten years. Such an instrument is expected to be able to directly image nearby, rocky planets orbiting M dwarfs in their HZs \citep{mazin19}. The expected end-goal IWA of the PSI is projected to be near $2\lambda/D$ for 10$^{-8}$ contrast in the J band \citep{sallum22}. This corresponds to an IWA of $\sim$ 0.017$^{\prime \prime}$, and would be capable of imaging GJ 251 c. Practical limits on the contrast capabilities of TMT and PSI will not be known until the instrument sees first light, though \cite{guyon18} provide some tools for estimating what the performance for a particular star may be. Even still, achievable contrasts depend on stellar temperature, magnitude, calibration stars, and a variety of other factors that are difficult to know before the instrument is tested. Consequently, the contrast upper limit shown in Figure \ref{fig:direct_imaging} is a useful tool to help assess the \textit{relative} feasibility of imaging nearby planets, but a more nuanced analysis is required to capture full feasibility. Using current predictive tools, we show that GJ 251 c might plausibly fall in the range of viable contrasts for PSI, though this requires some optimistic predictions for PSI wavefront control (this analysis is detailed further in Appendix B).

Due to its higher declination ($\delta$ = 33.26640801$^{\circ}$), GJ 251 can only be viewed via Southern hemisphere ELTs at relatively high airmass. Such observations introduce atmospheric effects that likely rule out sufficiently high contrast imaging from the Southern ELTs such as GMT and ESO ELT.

The planned Habitable Worlds Observatory (HWO) mission is a space based mission intent on imaging planets in the HZs of nearby stars \citep{decadal20}. The HWO is expected to reach contrasts up to $10^{-10}$ \citep{harada24}. Unfortunately, the $\sim$6 m primary mirror will likely limit the HWO's IWA to 0.065$^{\prime \prime}$, likely placing GJ 251 c out of reach, along with most HZ planets orbiting M dwarfs.

Direct imaging of an exoplanet allows astronomers to probe into the planetary atmosphere, potentially revealing a plethora of knowledge about the distant world. For example, the PSI will have the ability to detect such planetary features as cloud hazes and rotation rate \citep{crossfield14}, surface features \citep{karalidi12}, or even biomarkers using high-dispersion coronagraphy \citep{wang17}. 

Directly imaging the nearest planets is the only way to detect atmospheric features on nearby planets for the foreseeable future. Transmission spectroscopy \citep{benneke12} is an alternative method for retrieving atmospheric compositions of exoplanets, though it requires a planet to transit from Earth's perspective for observations to be obtained at all. This likely precludes any nearby HZ planets from study with transmission spectroscopy, with the exception of a small number of special systems (i.e.~TRAPPIST-1). \cite{hardegree-ullman23} estimate that no HZ planets transit inside of 5 pc, one or two might transit inside of 10 pc, and that 1-20 transit inside of 20 pc. All of these numbers are based upon an optimistic estimate for $\eta_{\oplus}$ ($>0.15$), the abundance of Earth-sized planets in the HZ of M dwarfs \citep{dressing15}, a poorly constrained quantity.

Direct imaging is likely to remain the primary method for studying the atmospheres of nearby exoplanets in the HZ, and GJ 251 c stands out as the best potentially terrestrial target for such studies after Proxima Centauri b \citep{angladescude16,faria22}, and the best in the celestial Northern Hemisphere. Furthermore, different stellar spectral types can have wildly different effects on the surface conditions of HZ planets \citep{shields14,shields16}, encouraging study of as many planets as possible.

\subsection{Climate Modeling}

To explore theoretical imaging spectra of GJ 251 c, we conduct climate simulations for four different atmospheric compositions; Earth-like, 10 bar CO$_2$, Titan-like, and hydrogen dominated mini-Neptune.  Table \ref{tab:climate_results} describes the atmospheric compositions, surface types and basic climate results.  The Earth-like composition is aligned with pre-industrial Earth conditions, but with O$_2$ and O$_3$ omitted due to lack of self-consistent photochemistry in our model that can accurately predict O$_3$ accumulation around an M-dwarf star.  The Titan-like composition is taken from \citet{horst2017titan} with an assumed haze production rate of 10$^{14}$ grams per Earth year, an approximate value taken from laboratory experiments of \citet{trainer2004,trainer2006organic}.  The mini-Neptune composition is taken from photochemical calculations of exoplanet K2-18 b performed by \citet{wogan2024jwst}.

\begin{deluxetable*}{ccccccc}
\label{tab:climate_results}
\tablecaption{Description of 3D climate simulations}
\tablehead{\colhead{Case} & \colhead{Atmospheric Composition} & \colhead{Pressure[bar]} & \colhead{Surface} & \colhead{TS[K]} & \colhead{TOA albedo} &  \colhead{OLR[$W/m^2$]}
}
\startdata
Earth-like & N$_2$, 285 ppm CO$_2$, 0.8 ppm CH$_4$, H$_2$O & 1 & ocean & 150.9 & 0.785 & 30.4 \\ 
$\mathrm{CO_2}$ Dominated & CO$_2$, H$_2$O & 10 & ocean & 319.7 & 0.144 & 117.2\\  
 Titan-like  & N$_2$, 5.65\% CH$_4$,  10 ppm C$_2$H$_6$, 10 ppb CO$_2$, H$_2$O, hazes & 1.5 & ocean & 158.5 & 0.322 & 95.6\\  
Mini-Neptune & H$_2$, 0.4\% CH$_4$, 0.01\% CO$_2$, 5\% H$_2$O & 10 & solid & 523.1 & 0.056 & 128.1
\enddata
\end{deluxetable*}

We use ExoCAM\footnote{https://github.com/storyofthewolf/ExoCAM} \citep{Wolf2021,wolf2022,neale:2010a}, a 3D climate model for terrestrial exoplanet atmospheres based on the Community Earth System Model (CESM) version 1.2.1\footnote{https://www.cesm.ucar.edu/models/cesm1.2/tags/} from the National Center for Atmospheric Research. Photochemical hazes are simulated in the Titan-like case using the Community Aerosol and Radiation Module for Atmospheres \citep[CARMA,][]{turco1979, toon:1988, bardeen2008}, which is fully coupled to ExoCAM.  Hazes are treated as fractal aggregates following from past CESM-CARMA modeling studies for early Earth \citep{wolf2010} and Titan \citep{larson2014, larson2015}. We assume an ocean surface for all cases except for the Mini-Neptune simulation, where we utilize a solid surface with an albedo of 0.2. The Earth-like and Titan-like cases, the ocean freezes and is treated as a solid ice surface. Albedo values are taken from \cite{Shields2013}. Geophysical properties of the planet were taken from Table \ref{tab:2p_posteriors}.

We find that GJ 251 c can maintain a habitable surface environment if it has a multibar CO$_2$ dominated atmosphere.  With a 10 bar CO$_2$ atmosphere and ocean surface as modeled here, the planet has a mean surface temperature of 319.7~K, an albedo of 0.144, and open ocean exists over its entire surface.  It is likely that significantly less CO$_2$ could still maintain habitable surface conditions for GJ 251 c.

Alternatively, an Earth-like composition has too few greenhouse gases to remain warm, has global ice coverage, and a cold global mean surface temperature of 150.9~K with an albedo of 0.785.   The Titan-like atmospheric case is also globally ice covered with a global mean surface temperature of 158.5~K and albedo of 0.322. The mini Neptune case is too hot for habitability, with a surface temperature at 10 bar of 523.1~K owing to the intense collision-induced-absorption (CIA) from its hydrogen dominated atmosphere.  The mini-Neptune world has a low albedo of 0.056 due to strong stellar absorption by CH$_4$, H$_2$O, and H$_2$-H$_2$ CIA. In all cases, cloud decks are thickest over the substellar point, typical of tidally locked planets around M-dwarf stars \citep{kopparapu2017}.  

\begin{figure*}
    \centering    \includegraphics[width=\textwidth]{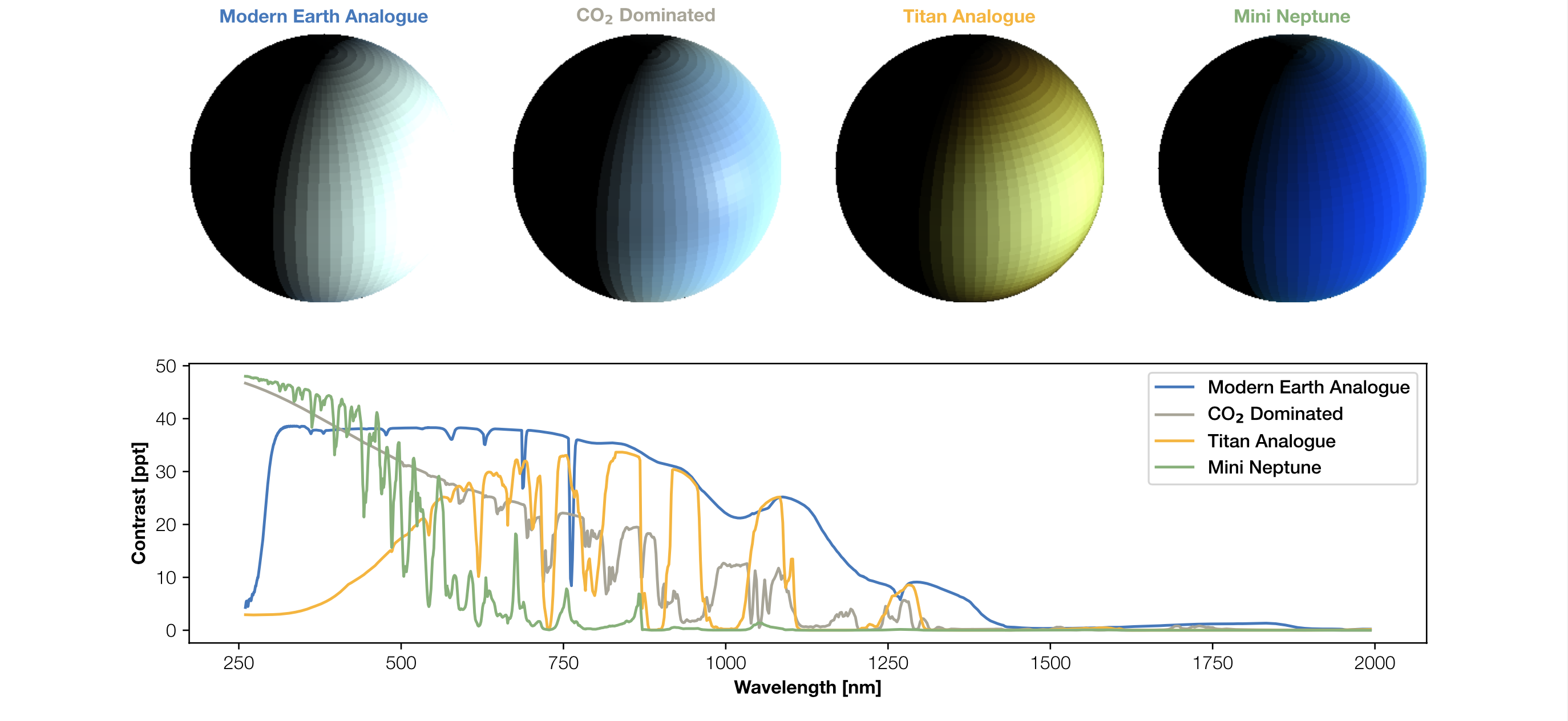}
    \caption{
Top panel: Visual representations of the four different climate states simulated for this study, with illuminated portions correlating with albedo. For these cases, the host-star's spectrum was set to a 5770 K blackbody to correspond to the Sun for the purpose of visual interpretation of the atmospheres.
Bottom panel: Contrasts of the different atmospheres as a function of wavelength. 
}
    \label{fig:climate_spectra}
\end{figure*}

We used the Planetary Spectrum Generator (PSG), an online tool designed to simulate radiative transfer in planetary atmospheres \citep{Villa2018}, to generate planetary spectra. For these simulations, the PSG was used (\url{https://psg.gsfc.nasa.gov/}) following the methods outlined in  \cite{kofman24}. The ExoCam multidimensional output is loaded into the GlobES module of PSG, which allows high-spatial-resolution simulations across the planet's sphere. Here, two types of simulations are done (Figure \ref{fig:climate_spectra}). In the first row of the Figure, spatially resolved RGB (red-green and blue) images of the four different climate states of the planet are shown, highlighting the locations of the clouds and the effects of the different processes dominating the visible (i.e.~Rayleigh scattering and absorption by hazes). The images are simulated at a resolution of 100 nm covering 400-500, 500-600 and 600-700 nm (blue, green and red respectively).

For the spectral simulations in the bottom panel, the spatial resolution is binned down by a factor of $4 \times 4$. The intensity units are in ‘contrast with respect to the host star’, which for these simulations is assumed to be a 3300 K blackbody emitting M-dwarf, with a radius of 0.25 R$_{\odot}$, the planet at a distance of 0.2 AU. The spectra below can be used as an indicator of the brightness of the planet, which in relative terms is approximately 300 times as bright as Earth around a G-star.

In terms of the composition of the atmosphere, it should be noted that the atmospheres are simplified, and it would be expected that more trace species ($\mathrm{CH_{4}}$, $\mathrm{C_{2}H_{6}}$) would be present in a real spectrum.

The modern Earth analogue is dominated by its familiar O$_{2}$ signatures (760 and 690 nm) and a strong drop-off in reflectivity below 300 nm due to O$_{3}$ and to a lesser extent $\mathrm{O_2}$. The broad features around 1000 and 1250 nm can be attributed to the reflectivity of the icy surface.
 
The $\mathrm{CO_{2}}$ dominated planet shows few strong features in the visible range. Rayleigh scattering is responsible for the increased flux towards shorter wavelengths, and the near IR absorption features between 600 and 1200 nm originate from water. Above this wavelength range $\mathrm{CO_{2}}$ dominates the absorbance, effectively leaving little flux to escape the planet. The produces the higher surface temperature and liquid  $\mathrm{H_{2}O}$  surface.
 
The reflection spectrum of the Titan analog planet is dominated by scattering and absorption from the hazes, Rayleigh scattering, and molecular absorption from  $\mathrm{CH_4}$.

Atmospheric simulations demonstrate the wide variety of possible surface conditions that might exist on GJ 251 c, while also highlighting the valuable information that might be gleaned from future atmospheric observations. Future spectra taken of GJ 251 c might conceivably give us a deep insight into a world very different from the Earth we know.

\subsection{Planet? Candidate Planet? Or False Positive?}

We conclude that the 54 day signal detected in the GJ 251 RV data is most likely planetary in origin, and call it a candidate planet, GJ 251 c. The difference between a candidate planet and a planet in literature is not well-defined, though we primarily choose this designation in light of our model comparisons (Table \ref{tab:modelcomparison}) only just exceeding the generally accepted BF 5 threshold. We would not be the first to claim a planet detection, only to find with the addition of more data that the signal was a false positive \citep{robertson14,robertson15,lubin21}. We compare GJ 251 c to a number of similar systems with previously detected false positive planets, and we highlight how GJ 251 c differs.

We first compare GJ 251 to the GJ 581 system. GJ 581 is very similar to GJ 251 in spectral type (M3V), distance from Earth (6.3 pc), and stellar rotation period \citep[130 $\pm$ 2 days;][]{robertson14}. Indeed, the system also shares some architectural similarities, with a known planet GJ 581 c (P$_{orb}$ = 12.91 days) similar to GJ 251 b (P$_{orb}$ = 14.2 days), and a claimed HZ planet orbiting at a period of 83 days \citep{udry07}, then later 66 days \citep{mayor09,vogt12}. The planet was eventually shown to be a false positive \citep{robertson14}.

\cite{robertson14}'s primary reasoning was a correlation between GJ 581 RVs and H$\alpha$ activity indicators, indicating that the RV data were contaminated with activity. This is especially damning in light of the fact that the claimed orbital period was almost exactly half of the stellar rotation period. This approach was more recently validated in \cite{stauffenberg24}. In the GJ 251 system, H$\alpha$ data are not available for most instruments, only CARMENES and NEID, so a directly analogous analysis is not possible. CARMENES and NEID H$\alpha$ data do show some correlation with RVs when calculated \citep[p$_{\mathrm{CARMENES}}$ = -0.28; p$_{\mathrm{NEID}}$ = -0.34; ][]{pearson1895}, though less than in the GJ 581 data. More importantly, this only indicates stellar activity contamination is present, and does not indicate timescales. H$\alpha$ periodogram analysis indicates that this correlation is present at longer timescales, not at any periods related to GJ 251 c. This is especially in contrast with GJ 581, where H$\alpha$ data did show activity on planet-related timescales. \cite{robertson14} used decorrelation to remove activity signals from RVs, though we use GPs, and they note a significant drop in LS power for GJ 581 d after decorrelation. We cannot easily perform the same test, as GP models are trained during inference and should not simply be subtracted from RVs to run new periodograms. We can compare two planet fits that utilized a GP, and ones that did not, and highlight that the recovered amplitude of planet c does not vary greatly between the cases, an analogous comparison to the decorrelated or non-decorrelated model in \cite{robertson14}.

We next compare GJ 251 to Alpha Centuari B b, a planet first identified by \cite{dumusque12}, but later shown to be a false positive \citep{rajpaul16}. Alpha Centauri B is a K1 V star in the Alpha Centauri system, the closest star system to Earth. The false positive planet had an amplitude of 0.5 m s$^{-1}$ and an orbital period of 3.2 days. \cite{rajpaul16} highlighted a number of features during analysis that suggested that the planet was not genuine, all of which do not apply to GJ 251 c. Alpha Centuari B b was modeled in four separate seasons, rather than utilizing all data simultaneously. We utilize all GJ 251 data simultaneously. \cite{dumusque12} used a series of sine waves to model the stellar activity, which as \cite{rajpaul16} point out, is a suboptimal model of the quasi-periodic variability that stellar activity signals typically generate. They recommend using a GP model, which we use for GJ 251. Finally, \cite{rajpaul16} found a significant power near the planetary period in the system's window function, which is not the case for GJ 251 c (Figure \ref{fig:window}).

%We next compare GJ 251 to Kapetyn's Star, another M dwarf close to earth \citep[d = 3.9 pc;][]{robertson15}. \cite{AE14} identified two super Earths orbiting Kapetyn's star, Kapetyn b and c. Kapetyn b was thought to orbit in the star's HZ, and had a similar orbital period to GJ 251 c (P$_{orb}$ = 48.6 days), though \cite{robertson15} would go on to rule out a planetary explanation for the signal. Their primary line of reasoning was a significant detection of the stellar rotation period in activity indicators at almost exactly three times the planet orbital period (142.9 $\pm$ 0.3 days). Furthermore, a 48 day signal appeared with high power in four activity indicators, though it was never the highest power signal. In contrast, a periodicity near 54 days never shows up strongly in the activity indicators. A small power near 54 days does appear in the system's CA IRT II triplet data (Figure \ref{fig:indicator_periodogram}), though it is far from the largest peak, and statistically insignificant, a notable difference from Kapetyn b.

Finally, we compare GJ 251 to Barnard's star, an old M Dwarf, the second closest star system to the Sun, and the closest in the celestial Northern Hemisphere \citep{bailer-jones21}. \cite{ribas18} identified a planet orbiting the star with a period of 233 days, though \cite{lubin21}, and later \cite{gonzalez24}, would eventually demonstrate that Barnard's Star b was a false positive, a particularly insidious case of aliasing. \cite{lubin21} show that the 233 day signal is a one year alias with the stellar rotation period. They also highlight that most of the planetary power comes from a particular region of RV data, and that during this region stellar activity seems to have spiked in the S$_{HK}$ values as well as in H$\alpha$. Furthermore,  these spikes occurred at the proposed planetary period. 

GJ 251 c, in contrast, is not created by any common aliases with the stellar rotation period, nor does the signal ever have significant power in any of the stellar activity indicators. An analysis of the window function of GJ 251 data (Figure \ref{fig:window}) further suggests that no common sampling timescales contribute strongly to the data, with the exception of a lunar cycle at 29 days. Our by-region and by-instrument analyses in Figure \ref{fig:season_results} show that no particular region of time, or instrument, is contributing all of the power to the detection of GJ 251 c. Additionally, \cite{lubin21} found that an activity-only model was significantly preferred to a fit with a single planet, or a single planet and stellar activity. On the other hand, the top model for GJ 251 c is a significant improvement over all 0 planet models in our analysis (Table \ref{tab:modelcomparison}), even those that utilize GPs.

\section{Summary}\label{sec:summary}

Using archival HIRES, CARMENES, and SPIRou RVs in combination with precise new HPF and NEID RVs, we explore the many periodic signals in GJ 251's time series. The high precision of NEID RVs allowed us to experiment with utilizing only the reddest orders in order to create a pseudo-NIR instrument. New chromatic GP kernels from \cite{cale21} allowed us to explore wavelength dependence, and a robust model comparison exploring more than 50 different models allowed us to explore GJ 251 more deeply than has ever been done before. We identify six signals that prominently appear in the data of GJ 251, and confirm the first as a planet (14 days), four as stellar activity related (68, 73, 122, and 133 days), and one as likely planetary (54 days). 

 GJ 251's proximity to Earth and its favorable spectral type make it a prime target for future direct imaging missions on next generation telescopes, and it is a particularly valuable Northern target.

Because GJ 251 c orbits in the HZ of a nearby star, it is likely one of the most promising places to search for biosignatures with next generation telescopes. We show using climate simulations that temperate surface conditions could exist on GJ 251 c, and we highlight spectral features that may be observed via direct imaging.

\section{Acknowledgements}

%Data
Data used in this work, and modified source code are permanently linked at \url{https://doi.org/10.5281/zenodo.17238980}.

%TESS guest investigator
This work was partially supported by NASA grant 80NSSC22K0120 to support Guest Investigator programs for TESS Cycle 4. 

%FINESST
This work was partially support by the Future Investigators in NASA Earth and Space Science and Technology (FINESST) program Grant No. 80NSSC22K1754.

%referees
We thank the referees for their thorough feedback on the manuscript. They have improved the quality of manuscript as a whole.

%WIYN Acknowledgement

Based on observations at Kitt Peak National Observatory, NSF’s NOIRLab (Prop. ID 2020B-0422; PI: A. Lin. Prop. ID 2021A-0385; PI: A. Lin. Prop. ID 2021B-0435; PI: S. Kanodia. Prop ID 2021B-0035; PI: S. Kanodia), managed by the Association of Universities for Research in Astronomy (AURA) under a cooperative agreement with the National Science Foundation. The authors are honored to be permitted
to conduct astronomical research on Iolkam Du’ag (Kitt
Peak), a mountain with particular significance to the
Tohono O’odham.

%NEID Acknowledgement

This paper contains data taken with the NEID instrument, which was funded by the NASA-NSF Exoplanet Observational Research (NN-EXPLORE) partnership and built by Pennsylvania State University. NEID is installed on the WIYN telescope, which is operated by the NSF's National Optical-Infrared Astronomy Research Laboratory (NOIRLab), and the NEID archive is operated by the NASA Exoplanet Science Institute at the California Institute of Technology. NN-EXPLORE is managed by the Jet Propulsion Laboratory, California Institute of Technology under contract with the National Aeronautics and Space Administration.

%Queue Observer Acknowledgement

We thank the NEID Queue Observers and WIYN Observing Associates for their skillful execution of our NEID observations.

%HET

The Hobby-Eberly Telescope (HET) is a joint project of the University of Texas at Austin, the Pennsylvania State University, Ludwig-Maximilians-Universität München, and Georg-August-Universität Göttingen. The HET is named in honor of its principal benefactors, William P. Hobby and Robert E. Eberly.

%HPF

These results are based on observations obtained with the Habitable-zone Planet Finder Spectrograph on the HET. The HPF team was supported by NSF grants AST-1006676, AST-1126413, AST-1310885, AST-1517592, AST-1310875, AST-1910954, AST-1907622, AST-1909506, AST-2108493, AST-2108512, AST-2108569, AST-2108801, ATI-2009889, ATI-2009982, ATI-2009554, and the NASA Astrobiology Institute (NNA09DA76A) in the pursuit of precision radial velocities in the NIR. The HPF team was also supported by the Heising-Simons Foundation via grant 2017-0494.

%Mauna Kea
The authors wish to recognize and acknowledge the very significant cultural role and reverence that the summit of Maunakea has always had within the Native Hawaiian community. We are most fortunate to have the opportunity to conduct observations from this mountain.

%CHAMPS
This work was partially completed as part of NASA’s CHAMPs team, supported by NASA, United States under Grant No. 80NSSC21K0905 issued through the Interdisciplinary Consortia for Astrobiology Research (ICAR) program.

%CEHW
The Center for Exoplanets and Habitable Worlds is supported by the Pennsylvania State University and the Eberly College of Science.

%KOA

This research has made use of the Keck Observatory Archive (KOA), which is operated by the W. M. Keck Observatory and the NASA Exoplanet Science Institute (NExScI), under contract with the National Aeronautics and Space Administration.

%Gaia DR3

This work has made use of data from the European Space Agency (ESA) mission
{\it Gaia} (\url{https://www.cosmos.esa.int/gaia}), processed by the {\it Gaia}
Data Processing and Analysis Consortium (DPAC,
\url{https://www.cosmos.esa.int/web/gaia/dpac/consortium}). Funding for the DPAC
has been provided by national institutions, in particular the institutions
participating in the {\it Gaia} Multilateral Agreement.

%   JPL
The research was carried out, in part, at the Jet Propulsion Laboratory, California Institute of Technology, under a contract with the National Aeronautics and Space Administration (80NM0018D0004).

%TESS Acknowledgement
This paper includes data collected by the TESS mission. Funding for the TESS mission is provided by the NASA's Science Mission Directorate.

VK is supported by the GSFC Sellers Exoplanet Environments Collaboration (SEEC) and the Exoplanets Spectroscopy Technologies (ExoSpec), which are part of the NASA Astrophysics Science Division’s Internal Scientist Funding Model.

The Center for Exoplanets and Habitable Worlds is supported by Penn State and its Eberly College of Science.

\facilities{\gaia{}, WIYN (NEID), HET (HPF), \tess{}, Exoplanet Archive}
\software{
\texttt{astropy} \citep{astropy18},
\texttt{barycorrpy} \citep{kanodia18b}, 
\texttt{ipython} \citep{ipython07},
\texttt{lightkurve} \citep{lightkurve},
\texttt{matplotlib} \citep{Hunter07},
\texttt{numpy} \citep{harris20},
\texttt{pandas} \citep{reback2020pandas,mckinney-proc-scipy-2010},
\texttt{RadVel}\citep{fulton18},
\texttt{scipy} \citep{2020SciPy-NMeth},
\texttt{SERVAL} \citep{zechmeister18},
}

\bibliography{bibliography}

\appendix

\onecolumngrid

\renewcommand{\thefigure}{A\arabic{figure}}
\setcounter{figure}{0}
\renewcommand{\thetable}{A\arabic{table}}
\setcounter{table}{0}

\section{Additional Plots and Figures}
\label{sec:appendixA}

\begin{figure}[b]
    \centering
    \includegraphics[width=0.5\textwidth]{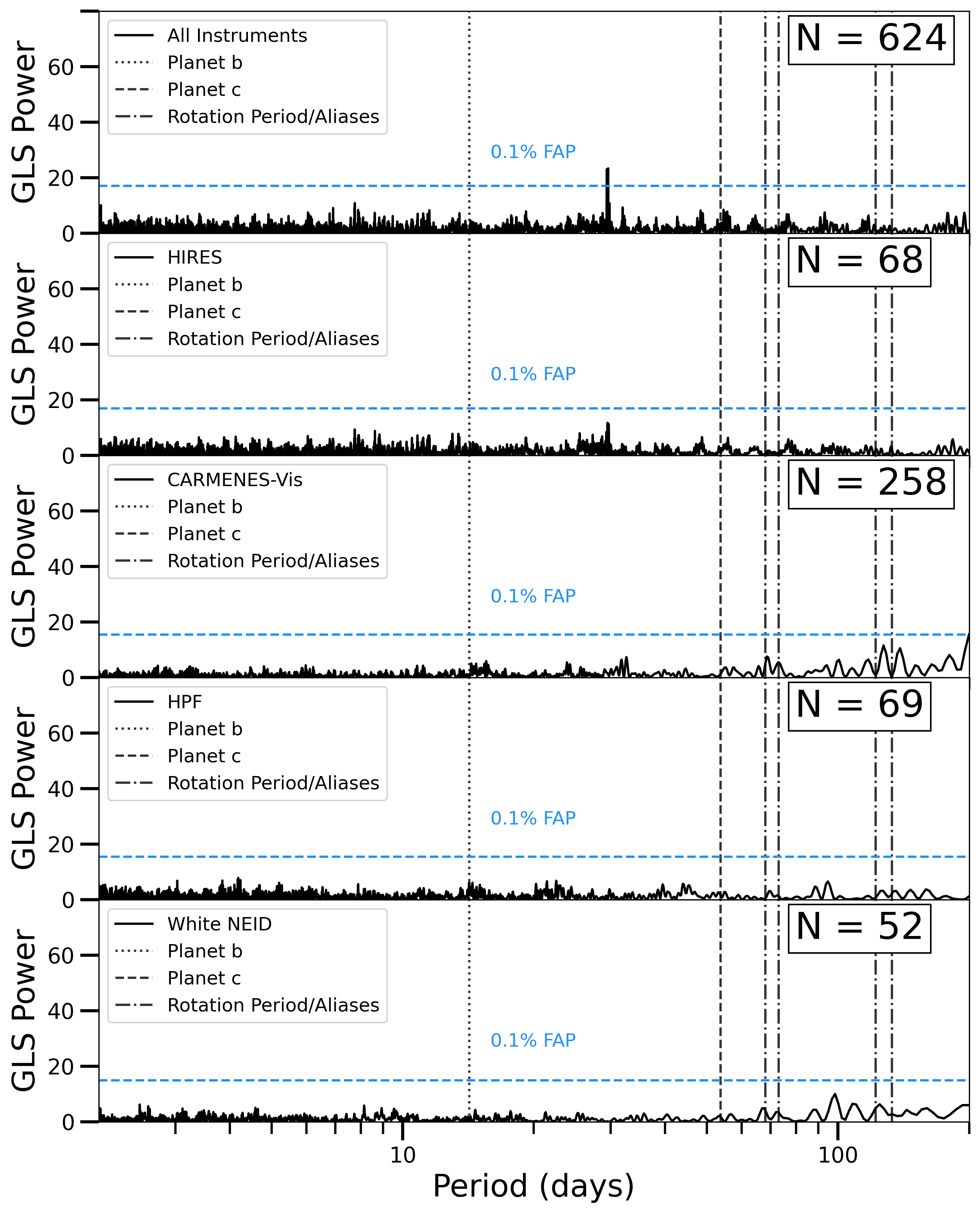}
    \caption{Window function of our data for each instrument. Only a periodicity near thirty days, most likely a lunar cycle, seems to be caused by our sampling.}
    \label{fig:window}
\end{figure}

We include a number of additional figures that readers may find enlightening. First, we include a periodogram of the window function of our GJ 251 data in Figure \ref{fig:window}. The window function can be used to highlight timescales that our observational sampling may have poorly constrained. It can also emphasize timescales that may be artificially inflated in power due only to our sampling.

\begin{figure*}
   \centering
    \includegraphics[width=\textwidth]{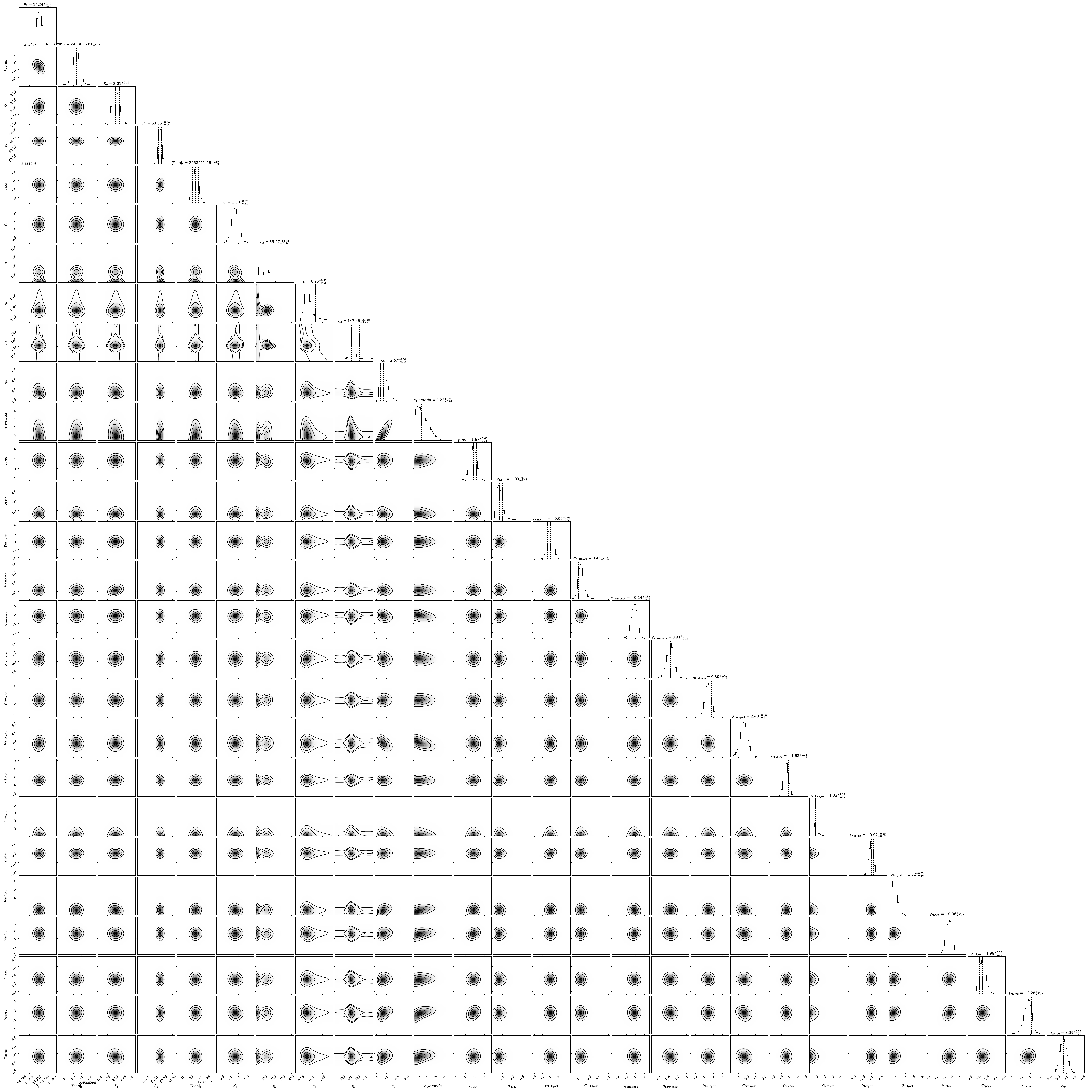}
    \caption{We show a corner plot of our best fit posteriors. Results are generally well behaved and normally distributed.}
   \label{fig:corner}
\end{figure*}

We next include a corner plot showing the posteriors of our preferred MCMC fit in Figure \ref{fig:corner}.

\begin{figure*}
    \centering
    \includegraphics[width=\textwidth]{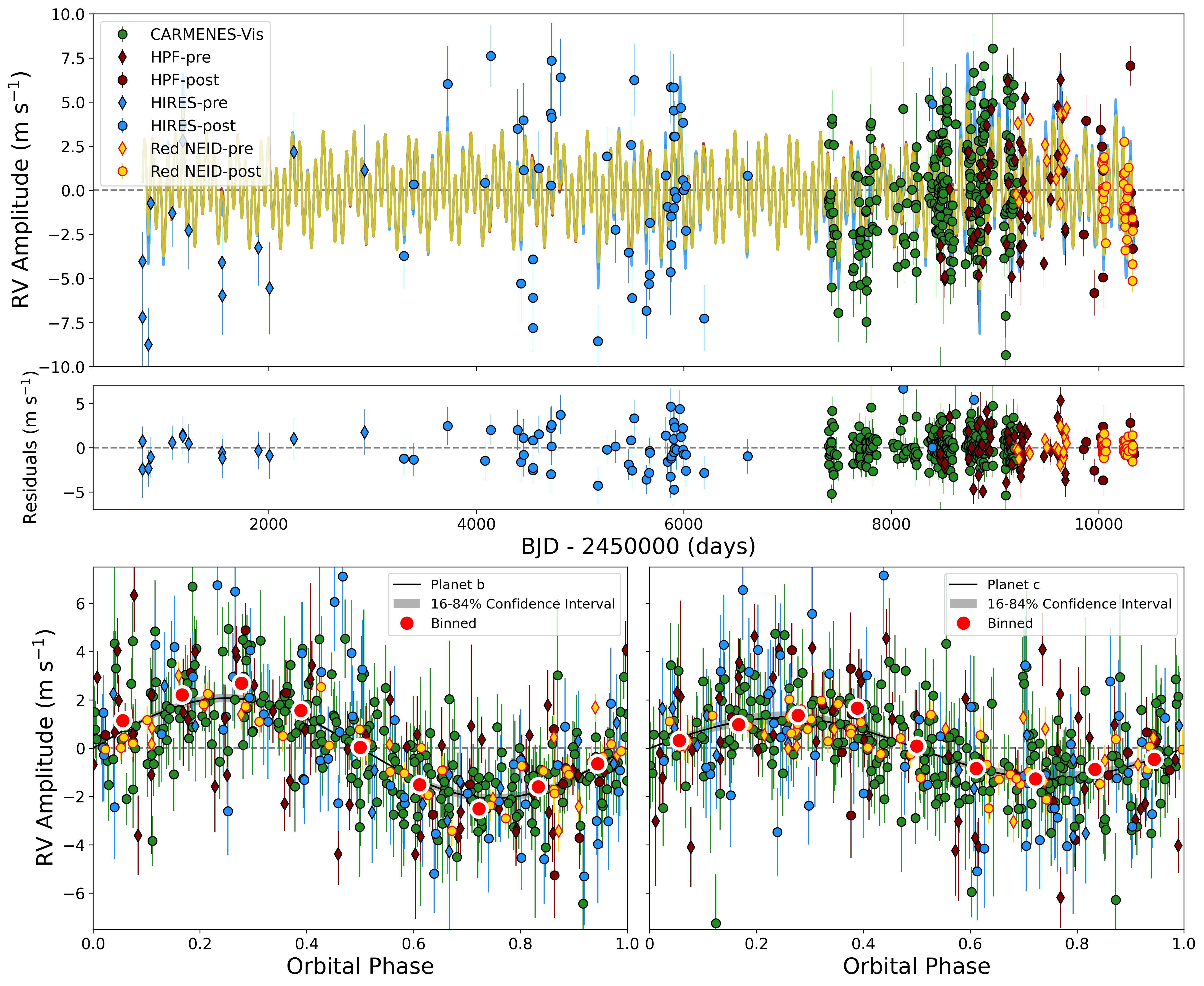}
    \caption{RV plot of our best Red NEID fit.}
    \label{fig:red_rv_plot}
\end{figure*}

Finally, we include a plot of the Red NEID RV results in Figure \ref{fig:red_rv_plot}.

\clearpage

\section{Contrast Limits on Thirty Meter Telescopes}

Much of the significance of the discovery of the planet candidate, GJ 251 c, is its status as a plausibly terrestrial HZ planet that might possibly be directly imaged on thirty meter class telescopes. The expected inner working angle of TMT will likely not preclude GJ 251 c from a future study, though instrumental contrast limitations may.

\begin{figure*}
    \centering
    \includegraphics[width=\textwidth]{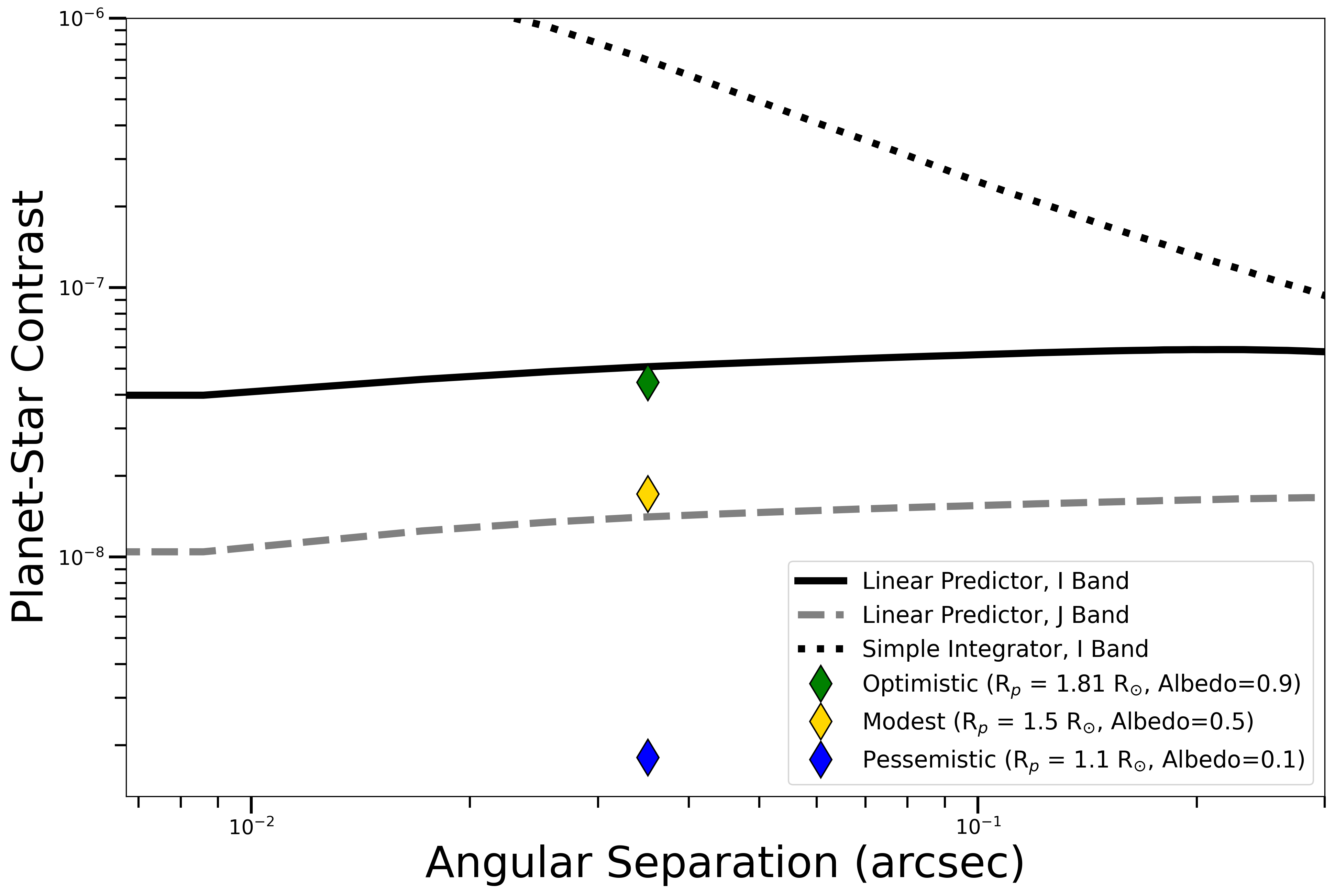}
    \caption{A variety of possible contrast limits for the TMT PSI are plotted as solid, dashed, and dotted lines. Calculated contrasts for GJ 251 c are shown for a variety of planet radius and albedo values. GJ 251 c will test the limitations of PSI, but there are a number of scenarios where its direct imaging will be possible.}
    \label{fig:contrast_limits}
\end{figure*}

In Figure \ref{fig:contrast_limits}, we use PSI-Sim\footnote{https://github.com/planetarysystemsimager/psisim} to show a few possible performance scenarios for TMT. We show optimistic scenarios with high (factor of 100) instrument performance gain. We plot limits based on current expected $I$-band imaging, as well as those in $J$-band, where GJ 251 is brighter. We show simple integration predictions as well as a linear predictor \citep[i.e.~predictive control;][]{guyon17}. We additionally plot a few estimated contrast values for GJ 251 assuming optimistic (R$_{p}$ = 1.81 R$_{\oplus}$, albedo=0.9), modest (R$_{p}$ = 1.5 R$_{\oplus}$, albedo=0.5), and pessimistic (R$_{p}$ = 1.1 R$_{\oplus}$, albedo=0.1) planet-star contrasts.

This analysis highlights the challenge of direct detection for any temperate, terrestrial exoplanet. GJ 251 c will likely only be imageable if the planet has a large albedo or radius, and will additionally depend on PSI performing very well. If the wavefront sensor can be moved to a bandpass in which GJ 251 is brighter (e.g. $J$-band), imaging may be possible even in the modest case. Even when considering these factors, GJ 251 c stands out among known systems as a highly plausible candidate.

\end{document}